\documentclass[british,useAMS,usenatbib]{mn2e}
\usepackage{mathptmx}
\usepackage[T1]{fontenc}
\usepackage[latin9]{inputenc}
\usepackage{graphicx}
\usepackage{verbatim}
\usepackage[authoryear]{natbib}

\providecommand{\LyX}{L\kern-.1667em\lower.25em\hbox{Y}\kern-.125emX\@}
\makeatletter

\@ifundefined{definecolor}
  {\usepackage{color}}{}

\usepackage{babel}
\makeatother

\begin{document}
\title[Where do long-period comets come from?]{Where do long-period comets come from?
\newline Moving through Jupiter--Saturn barrier.}

\author[P.A. Dybczy\'{n}ski \& M. Kr\'olikowska]{Piotr A. Dybczy\'{n}ski$^{1}$
\thanks{E-mail: dybol@amu.edu.pl}
\& Ma\l{}gorzata Kr\'olikowska$^2$
\thanks{E-mail:mkr@cbk.waw.pl}\\
$^1$Astronomical Observatory Institute,
A.Mickiewicz Univ., S\l{}oneczna 36, 60-286 Pozna\'{n}, Poland;\\
$^2$Space Research Centre of the Polish Academy of Sciences,
Bartycka 18A, 00-716 Warsaw, Poland.}

\date{}

\pagerange{\pageref{firstpage}--\pageref{lastpage}} \pubyear{2011}
\maketitle

\label{firstpage} 

\begin{abstract}
Past and future dynamical evolution of all 64 long period comets having
$1/a_{{\rm ori}}$< 1$\times10^{-4}$\,au$^{-1}$ and $q_{{\rm osc}}>3.0$\,au
and discovered after 1970 is studied. For all of them we obtained
a new, homogeneous set of osculating orbits, including 15 orbits with
detected non-gravitational parameters. The non-gravitational
effects for eleven of these 15 comets have been determined for the
first time. This means that more than 50\% of all comets with perihelion
distances between 3 and 4 au and discovered after 1970 show detectable
deviations from a purely gravitational motion. Each comet was then
replaced with the swarm of 5001 virtual comets well representing observations.
These swarms were propagated numerically back and forth up to the
250 au heliocentric distance, constituting sets of \emph{original}
and \emph{future} orbits together with their uncertainties. This allowed
us to show that the $1/a_{{\rm ori}}$ distribution is significantly
different in shape as well as the maximum position when including
non-gravitational orbits. Next we followed the dynamical evolution
under the Galactic tides for one orbital revolution to the past and
future, obtaining orbital elements at previous/next perihelion passages.
We obtained a clear dependence of the last revolution change in perihelion
distance with respect to the $1/a_{{\rm ori}}$, what confirmed theoretical
expectations.

Basing on these results we discuss the possibility of discriminating
between dynamically new and old comets with the aid of their previous
perihelion distance. We show that about 50 per cent of all investigated
comets have their previous perihelion distance below the 15\,au limit.
This resulted in classifying 31 comets as \emph{dynamically new},
26 as \emph{dynamically old} and 7 having unclear status. We
showed that this classification seems to be immune against to perturbations
from all known stars. However discoveries of new, strong stellar perturbers,
while rather improbable, may change the situation.

We also present several examples of cometary motion through the Jupiter-Saturn
barrier, some of them with the previous perihelion distance smaller
than the observed one. New interpretations of the long period comets
source pathways are also discussed in the light of suggestions of
\citet{kaib-quinn:2009}. 
\end{abstract}


\section{Introduction}

\label{sec:Introduction}

As Martin Duncan~\citeyearpar{duncan:2009} recently pointed out,
the forthcoming deep, wide-field observational surveys (both ground-based
and space-based) soon be providing completely new amount of data on
very large perihelion distance comets, far behind Saturn. These may
extend our possibilities to investigate the source regions of the
long period comets (LPCs). But up to the date, the only observational
constrains on the LPCs source is the detailed analysis of their observations,
ranging up to 10\,au or so, determining their orbits and studying
their past dynamical evolution. Continuing our effort in this field
(see \citealp{kroli-dyb:2010}, hereafter Paper I) we analysed the
sample of next 64~comets from the so called Oort spike, all of them
having the osculating perihelion distance greater than 3\,au, wherein
four comets with $q_{{\rm osc}}>3.0$\,au and determinable non-gravitational
(here and after NG) orbits were taken from Paper I. The detailed
description of our sample can be found in Section \ref{sec:Sample-description}.
In short, the reason for such a selection was to deal with comets
with negligible NG forces due to their large heliocentric distance.
However, during the data analysis we attempted to determine the NG
forces parameters and it appeared that for another 11 comets it was
reasonable to use such a model. Thus, including four comets from Paper
I we have in total 15~large perihelion distance comets with NG parameters
obtained and used in the original/future orbit determinations, details
can be found in Sections~\ref{sub:Non-gravitational-forces-detection} -- \ref{sub:NG-and-GR-models}.
In this paper we again ask for the source region of the Oort spike
comets. To this purpose we carefully analysed past and future dynamical
evolution of 64 comets under planetary and Galactic perturbations,
see Section \ref{sec:Prev-and-next-perihelion-passages}. This, additionally, makes
an opportunity to observe how the mechanism widely called \emph{Jupiter-Saturn
barrier} works in practice, see Sections \ref{sub:On-Jupiter-Saturn-barrier}
and \ref{sec:Prev-and-next-perihelion-passages}. The widely disputed
problem of discriminating between dynamically (and physically) \emph{new}
and \emph{old} comets is revisited in Section \ref{sec:New-and-old}
including a discussion of our results in the light of the newly proposed alternative cometary origin scenario \citep{kaib-quinn:2009}. Finall discussion and conclusions are presented in Section~\ref{sec:Conclusions}.

\begin{figure*}
\includegraphics[width=5.5cm,angle=-90]{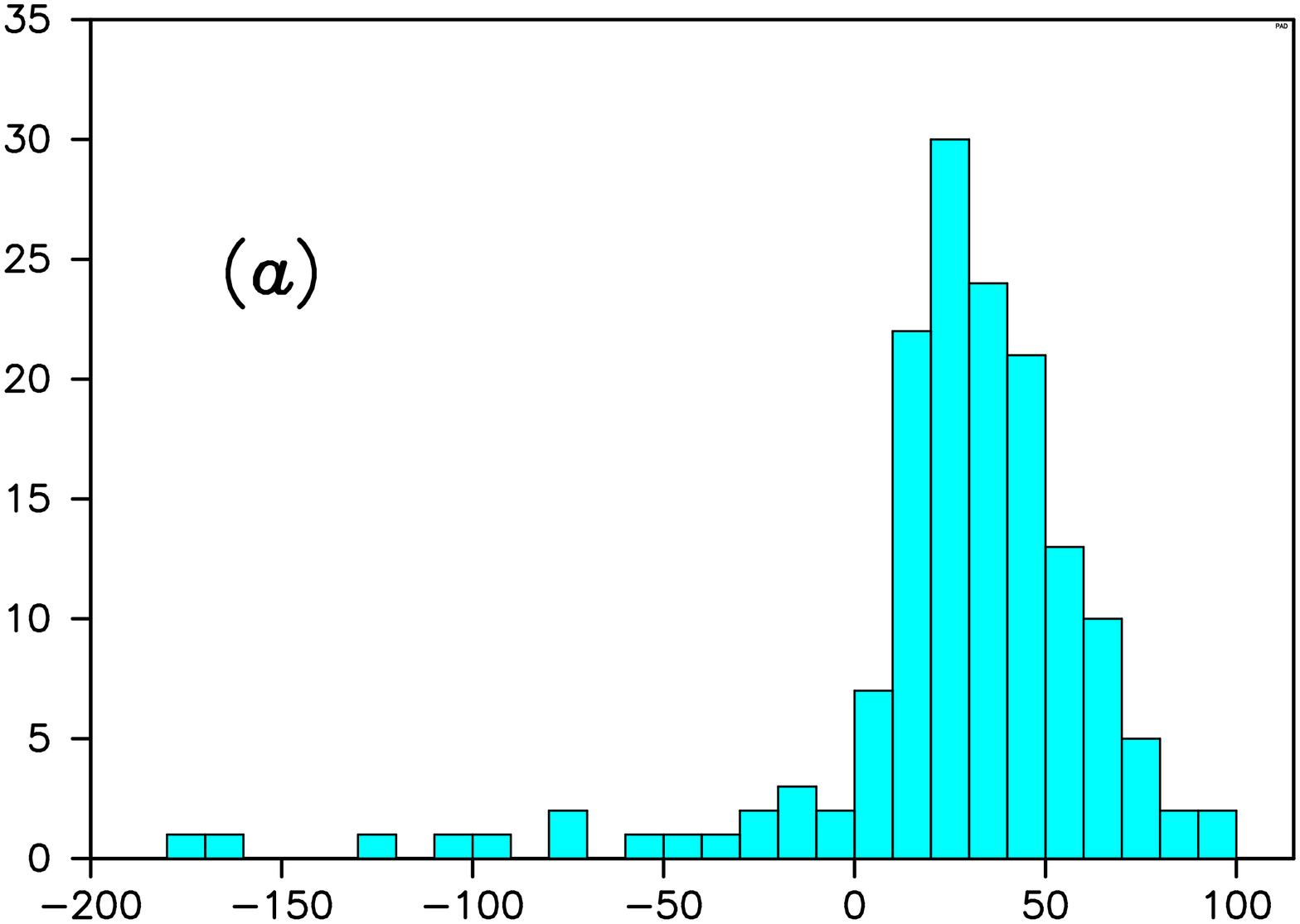}\hspace{1cm} \includegraphics[width=5.5cm,angle=-90]{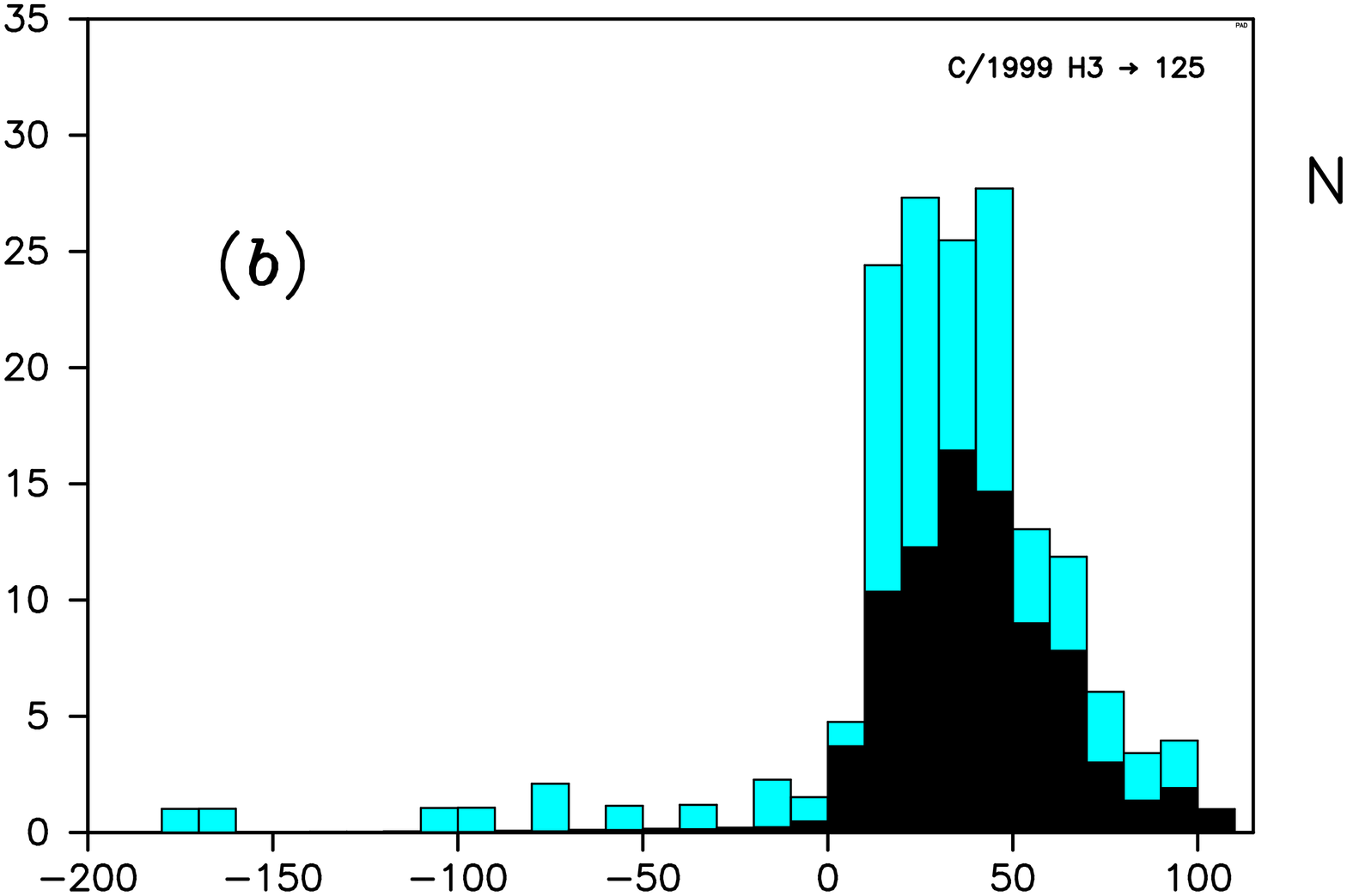}
\caption{\label{fig:1/a-for-all} Distribution of $1/a_{{\rm ori}}\times10^{6}$\,au$^{-1}$
for all LP comets with class 1 orbits and $1/a_{{\rm ori}}<10^{-4}$\,au$^{-1}$.
(a): 153 comets with $1/a_{{\rm ori}}$ taken directly from the MWC08 (splitting
product C/1996 J1-A omitted) , (b): 165 Oort spike comets; black part
of the histogram denotes 86~comets with orbits determined in this
paper and Paper I. We increased the MWC08 sample by 12 comets and
determined the NG orbits for 37 comets.}
\end{figure*}

\subsection{On the Jupiter-Saturn barrier}

\label{sub:On-Jupiter-Saturn-barrier}

The concept of Jupiter-Saturn barrier can be traced back by 30 years
or so. In 1981 \citeauthor{fernandez:1981} presented the dependence
of the comet energy changes due to the planetary perturbation on the
perihelion distance. It appeared, that for perihelion distances smaller
than 15\,au this perturbation is comparable with LP comet binding
energy.

Later on, as a result of some numerical simulations \citet{weissman:1985b}
stated that 65 per cent of comets coming closer to the Sun than Saturn
(and 94 per cent for Jupiter) will be ejected from the Solar System
as a result of planetary perturbations.

These results of planetary perturbations investigations were then
combined with the model of Galactic disk tides. \citet{matese-whitman:1989}
showed that it is possible to calculate the minimum cometary semimajor
axis, for which Galactic perturbations can decrease cometary perihelion
distance from above the strong planetary perturbation border down
below the observability limit in one orbital period. Using 15\,au
as the former and 5\,au as the later they obtained the minimum semimajor
axis to be equal 20\,000\,au. Investigating Galactic perturbations
in a way somewhat similar to ours, \citet{yabushita:1989} obtained
a very strong dependence of the perihelion distance reduction by Galactic
tides on the cometary semimajor axis, namely $\Delta q\sim a^{6.3\pm 0.2}$.
His conclusion was that 25\,000\,au is the most probable semimajor
axis of LP~comets arriving at the vicinity of the Sun for the first
time. Finally, in a review paper, \citet{fernandez:1994} devoted
a separate section to the description of the Jupiter-Saturn barrier
mechanism. This was also discussed by many authors, see for example
\citet{festou-r-w:1993b,levison-d-d:2001,dones-w-l-d:2004,fernandez_book:2005,morby:2005}.

This effect can be described in short as follows: the Galactic perturbations
can result in a continuous cometary perihelion drift towards the Sun.
The rate of this drift strongly depends on cometary semimajor axis,
which is almost constant here. If we agree that the perihelion distance
smaller than 10--15\,au results in the ejection from the Solar System
by perturbations from Jupiter and Saturn the only possibility for
a LP comets from the Oort cloud to become observable is to have their
perihelion distance reduced from above 10--15\,au down below the
observability limit, say 3--5\,au. Basing on the Galactic tide model
it is possible to calculate the minimum necessary semimajor axis to
accomplish this.

Such a calculation can be performed in several slightly different
formalisms but only two factors can change the result significantly:
the amplitude of the necessary perihelion distance reduction and the
assumed Galactic disk matter density $\rho$. In almost all papers
the authors demand the perihelion reduction by \textasciitilde{}10\,au
(from above 10--15\,au to below 3--5\,au) but they use different
disk matter densities. In earlier papers (\citealp{matese-whitman:1989,festou-r-w:1993b})
they used $\rho=0.185$ solar masses per cubic parsec, later (see for
example: \citealp{wiegert-tre:1999,morby:2005}) $\rho=0.150$ was
used. In all recent papers the value of $\rho=0.100$ solar masses
per cubic parsec is used, see \citet{levison-d-d:2001} for a supporting
discussion of this density value.

As a consequence, while previously obtained results ($a>\sim20\,000$\,au)
were roughly in line with observations, now the situation is more
complicated. In a review by \citet{dones-w-l-d:2004} one can read:
'If we assume that a comet must come within 3\,au of the Sun to become
active and thus observable, $\Delta q$ must be at least $\sim$10\,au
\textendash{} 3\,au $=$ 7\,au. It can be shown that, because of
the steep dependence of $\Delta q$ on $a$, this condition implies
that $a>28\,000$\,au'. But there are many observed Oort spike comets
with much smaller semimajor axes!

There is also an additional aspect of the Jupiter-Saturn barrier.
The resulting semimajor axis threshold value is often used as a definition
of the border between the outer (observable) and inner (unobservable)
parts of the Oort cloud. With the current value ($a>28\,000$~au)
this seems to be problematic.

Recently \citet{kaib-quinn:2009} revisited the Jupiter-Saturn barrier
problem and give several interesting new conclusions. We will discuss
them in Section \ref{sub:New-interpretations}.

In what follows we present some examples showing how different from
the above theory is the behaviour of many observed Oort spike comets.

\begin{center}
\begin{table*}
\caption{\label{tab:a_original_15}Original and future semimajor axes derived
from pure gravitational nominal solutions (columns~3--4) and NG~nominal
solutions (columns~5--6) for 15 large perihelion distance comets
with detectable NG~effects; the number of NG~parameters determined
for NG~solutions is given in the column~11. The rms's and number
of residuals are given in the columns 7--8 and 9--10, respectively.
The numbering in column 1 is consistent with the chronology of the
discovery of all 64 considered comets. Solutions for comets C/1997~BA$_{6}$,
C/1997~J2, C/1999~Y1 and C/2000~SV$_{74}$ are taken from Paper~I. }

{\setlength{\tabcolsep}{4.2pt} \begin{tabular}{@{}rlr@{$\pm$}rr@{$\pm$}rr@{$\pm$}rr@{$\pm$}rccrrr@{}}
\hline 
\#  & { Name }  & \multicolumn{4}{c}{{ gravitational solutions}} & \multicolumn{4}{c}{{ NG~solutions}} & { rms$_{{\rm GR}}$ }  & { rms$_{{\rm NG}}$ }  & \multicolumn{3}{c}{{ number of}}\tabularnewline
 &  & \multicolumn{2}{c}{{ $1/a_{{\rm ori}}$}} & \multicolumn{2}{c}{{ $1/a_{{\rm fut}}$}} & \multicolumn{2}{c}{{ $1/a_{{\rm ori}}$}} & \multicolumn{2}{c}{{ $1/a_{{\rm fut}}$}} &  &  & { res. }  & { res. }  & { NG}\tabularnewline
 &  & \multicolumn{4}{c}{{ i n ~~~u n i t s ~~~of ~~~$10^{-6}$\,au$^{-1}$}} & \multicolumn{4}{c}{{ i n ~~~u n i t s ~~~of ~~~$10^{-6}$\,au$^{-1}$}} & { \farcs }  & { \farcs }  & { GR }  & { NG }  & { par. }\tabularnewline
\hline 
{ 1 }  & { 2 }  & \multicolumn{2}{c}{{ 3 }} & \multicolumn{2}{c}{{ 4 }} & \multicolumn{2}{c}{{ 5 }} & \multicolumn{2}{c}{{ 6 }} & { 7 }  & { 8 }  & { 9 }  & { 10 }  & { 11 }\tabularnewline
\hline 
3  & { C/1974 F1 Lovas }  & { +36.82}  & { ~2.50}  & { ~+518.11}  & { ~2.50}  & { +40.48}  & { ~5.00}  
& {+522.07}  & { 6.81 }  & { 1.09 }  & { 1.08 }  & { 273}  & { 273}  & { 2}\tabularnewline
10  & { C/1980 E1 Bowell }  & { +27.52}  & { ~1.24}  & {-16014.}  & { ~~1.}  & { +53.35}  & { ~3.87}  & { -16011.}  & { ~4. }  & { 1.17 }  & { 1.06 }  & { 388}  & { 387}  & { 2}\tabularnewline
11  & { C/1983 O1 Cernis }  & { +47.79}  & { ~1.58}  & { ~-191.39}  & { ~1.58}  & { +60.83}  & { 36.12}  & { -186.95}  & { 2.08}  & { 1.14 }  & { 1.11 }  & { 461}  & { 461}  & { 2}\tabularnewline
12  & { C/1984 W2 Hartley }  & { +13.89}  & { 22.32}  & { ~~-27.92}  & { ~9.62}  & { +20.25}  & { ~8.56}  & { ~-31.55}  & { 8.57}  & { 1.89 }  & { 1.87 }  & { 107}  & { 107}  & { 1}\tabularnewline
21  & { C/1997 BA$_{6}$ Spacewatch}  & { ~+1.49}  & { ~0.35}  & { ~+371.77}  & { ~0.35}  & { +31.83}  & { ~1.15}  & { +402.48}  & { 1.72}  & { 0.74 }  & { 0.67 }  & { 1054}  & { 1054}  & { 3}\tabularnewline
22  & { C/1997 J2 Meunier-Dupouy}  & { +38.83}  & { ~0.57}  & { ~~~-2.72}  & { ~0.57}  & { +44.64}  & { ~0.88}  & { ~+14.72}  & { 0.91}  & { 0.67 }  & { 0.53 }  & { 2881}  & { 2863}  & { 2}\tabularnewline
25  & { C/1999 H3 LINEAR }  & { ~+65.99}  & { ~0.71}  & { ~~~-8.89}  & { ~0.71}  & {+124.66}  & { ~3.88}  & { ~-10.03}  & { 1.05}  & { 0.70 }  & { 0.51 }  & {1739}  & {1722}  & { 3}\tabularnewline
32  & { C/1999 Y1 LINEAR}  & { +42.92}  & { ~0.88}  & { ~+350.42}  & { ~0.88}  & { +47.35}  & { ~0.94}  & { +345.69}  & { 1.46}  & { 0.61 }  & { 0.48 }  & { 1747}  & { 1749}  & { 3}\tabularnewline
34  & { C/2000 CT$_{54}$ LINEAR }  & { +38.75}  & { ~1.35}  & { ~+557.96}  & { ~1.35}  & { +72.86}  & { ~3.06}  & { +585.28}  & { 2.53}  & { 0.90 }  & { 0.75 }  & { 418}  & {417}  & { 2}\tabularnewline
37  & { C/2000 SV$_{74}$ LINEAR}  & { +50.23}  & { ~0.40}  & { ~~-85.55}  & { ~0.40}  & { +92.31}  & { ~0.85}  & { ~-54.88}  & { 0.60}  & { 1.11 }  & { 0.71 }  & { 4389}  & { 4349}  & { 3}\tabularnewline
47  & { C/2002 R3 LONEOS }  & { +38.98}  & { ~0.57}  & { ~~~+9.88}  & { ~0.57}  & { +48.16}  & { ~3.07}  & { ~~+3.86}  & { 6.33}  & { 0.55 }  & { 0.52 }  & {2533}  & {2530}  & { 3}\tabularnewline
54  & { C/2005 B1 Christensen }  & { ~+7.14}  & { ~0.33}  & { ~+234.38}  & { ~0.33}  & { ~+3.99}  & { 0.61}  & { +239.09}  & { 0.47}  & { 0.45 }  & { 0.43 }  & { 2991}  & { 2985}  & { 3}\tabularnewline
55  & { C/2005 EL$_{173}$ LONEOS }  & { +45.70}  & { ~0.47}  & { ~~-35.29}  & { ~0.47}  & { +44.81}  & { 0.99}  & { ~-19.12}  & { 0.91}  & { 0.47 }  & { 0.36 }  & { 631}  & { 632}  & { 2}\tabularnewline
57  & { C/2005 K1 Skiff}  & { ~+7.66}  & { ~1.22}  & { ~~-79.34}  & { ~1.22}  & { +11.01}  & { ~2.75}  & { ~-82.36}  & { 3.35}  & { 0.52 }  & { 0.52 }  & {1257}  & {1254}  & { 2}\tabularnewline
61  & { C/2006 S2 LINEAR }  & { +73.92}  & { ~2.75}  & { ~~-30.83}  & { ~2.75}  & { +72.52}  & { ~8.14}  & { ~~-10.77}  & {18.00}  & { 0.45 }  & { 0.44 }  & { 346}  & { 346}  & { 2}\tabularnewline
\hline
\end{tabular}} 
\end{table*}

\par\end{center}

\section{The sample of large perihelion LP~comets}

\label{sec:Sample-description}

The latest published Catalogue of Cometary Orbits (\citealp{marsden-cat:2008},
hereafter MWC08) includes 154~Oort spike and hyperbolic comets ($1/a_{{\rm ori}}<10^{-4}$\,au$^{-1}$)
with orbits determined with the highest precision, i.e. with the quality
class 1, according to the classification introduced by \citet{mar-sek-eve:1978}.
Their $1/a_{{\rm ori}}$ distribution, based on catalogue data, is
shown in Fig.\ref{fig:1/a-for-all}(a). In the present investigation
we chose to study a sample of the Oort spike comets with large perihelion
distances, $q\geq3.0$\,au. The MWC08 contains 76~such comets. According
to the availability of astrometric data we restrict this sample to
comets discovered after 1970. Additionally, to have orbits definitely
determined we did not analyse three still potentially observable comets
in October 2010: C/2005~L3, C/2006~S3 and C/2007~D1. Thus, our
sample was reduced to 62~comets which orbits are of quality class~1
in MWC08. Recently, this sample has increased by two comets (C/2006~YC,
C/2007~Y1) that have orbits with the quality class~2 in MWC08 but
now their number and interval of observations have increased significantly
allowing for much more precise orbit determination. This makes the
total number of 64~LP comets studied in the present paper. The combined
$1/a_{{\rm ori}}$ distribution containing MWC08~comets superposed
and supplemented by our results is presented in Fig.~\ref{fig:1/a-for-all}\,(b).
One can observe the influence of our new results on the overall shape
of the histogram as well as position of $1/a_{{\rm ori}}$~-maximum.
The black part consists of all 86~comets investigated in Paper~I
and present research. It should be stressed here, that during the
orbit determination process we obtain $1/a_{{\rm ori}}$ value as
well as its uncertainty. This uncertainties are taken into account
when constructing the black part of the histogram depicted in Fig.~\ref{fig:1/a-for-all}\,(b).
It is evident, that incorporating NG effects in orbit determination
(where possible) moves the overall distribution towards smaller semimajor
axes.

For each comet from the sample we determined its osculating nominal
orbit (gravitational and NG if it was possible; see next section)
based on the data selected and weighted according to the methods described
in great detail in Paper~I (see also section~\ref{sub:Non-gravitational-forces-detection} therein). 
This allows
us to construct a homogeneous sample of cometary osculating orbits
which are the starting point for us to obtain the original and future
orbits with their uncertainties and then to study cometary dynamical
evolution under the Galactic tides.

Our set of large perihelion LP~comet orbits with $1/a_{{\rm ori}}<10^{-4}$\,au$^{-1}$
contains only one comet with slightly negative value of $1/a_{{\rm ori}}$,
namely C/1978~G2. The MWC08 attributes original hyperbolic orbits
to three more comets: C/1942~C2, C/2002~R3, C/2005~B1. The first
was discovered before 1970, therefore it was excluded from our sample.
In the present paper, orbits of two other comets are determined based
on a significantly longer interval of observations than in MWC08 and
now their original (gravitational) orbits derived by us are elliptical.
Later in this paper, it appears that comet C/1978~G2 may also be
of local origin. Therefore, we call this sample the large perihelion
distance Oort spike comets. For comets with small perihelion distances
we have an essentially different situation. In MWC08 nearly 20 comets
with $q<3.0$\,au have negative values of $1/a_{{\rm ori}}$
(see Fig.\ref{fig:1/a-for-all}a). It turns out that for these comets
important is the role of NG effects in the process of determining
their osculating orbits: most of small perihelion comets on hyperbolic
original orbits in the gravitational case have elliptical original
orbits when the NG forces are taken into account (see next section).

Observed perihelion distance and ecliptic inclination distributions
of all investigated comets are also shown and discussed in Section
\ref{sub:How-we-distinguish}.

\begin{table*}
\caption{\label{tab:past_motion_old} The past distributions of swarms of VCs
in terms of returning {[}R{]}, escaping {[}E{]}, including hyperbolic
{[}H{]} VC numbers for dynamically old comets. Aphelion and perihelion
distances are described either by a mean value for the normal distributions,
or three deciles at 10, 50 (i.e. median), and 90 per cent. In the
case of mixed swarm the mean values or deciles of $Q$ and $q$ are
given for the returning part of the VCs swarm, where the escape limit
of 120\,000\,au was generally used. The upper-a index in columns
5--7 means that this part of mixed swarm includes the nominal orbit.
For comparison we included the osculating perihelion distance in the
third column; in the fourth column - the galactic latitude of the perihelion direction is given.
The last two columns present the value of $1/a_{{\rm ori}}$
and the percentage of VCs, that we can call dynamically new, based
on previous $q$ statistics. Fifteen comets with NG~effects including
four objects from the Paper~I (No: 21,22,32,37) are indicated by
upper-NG index located behind the comet designation (column 2).}

\global\long\global\long\global\long\global\long\global\long\global\long\global\long\global\long\global\long\global\long\global\long\def\tabcolsep{0.6mm}

\centering{}\begin{tabular}{rlcccccccr@{$\pm$}rc}
\hline 
\#  & Comet  & $q_{{\rm osc}}$  & $b_{{\rm osc}}$  & \multicolumn{3}{c}{Number of VCs} & $Q_{{\rm prev}}$  & $q_{{\rm prev}}$  & \multicolumn{2}{c}{$1/a_{{\rm ori}}$} & \% of \tabularnewline
 &  & au  & deg  & {[}R{]}  & {[}E{]}  & {[}H{]}  & $10^{3}$au  & au  & \multicolumn{2}{c}{$10^{-6}$au$^{-1}$} & dyn. new \tabularnewline
 &  &  &  &  &  &  &  &  & \multicolumn{2}{c}{ } & 10\,au - 15\,au - 20\,au \tabularnewline
{[}1{]}  & {[}2{]}  & {[}3{]}  & {[}4{]}  & {[}5{]}  & {[}6{]}  & {[}7{]}  & {[}8{]}  & {[}9{]}  & \multicolumn{2}{c}{{[}10{]}} & {[}11{]} \tabularnewline
\hline 
{ 1 }  & C/1972 L1  & 4.28  & 40.3  & 5001  & 0  & 0  & 33.9 - 39.0 - 46.5  & 4.06 - 4.18 - 5.08  & 51.1  & 6.2  & 0.5 - 0.2 - 0.1 \tabularnewline
{ 2 }  & C/1973 W1  & 3.84  & -48.7  & 5001  & 0  & 0  & 22.7 - 27.5 - 35.2  & 4.14 - 4.47 - 5.53  & 72.6  & 12.3  & 0.7 - 0.4 - 0.2 \tabularnewline
{ 3 }  & {C/1974 F1$^{{\rm NG}}$}  & 3.01  & -21.7  & 5001  & 0  & 0  & 42.7 - 49.4 - 58.6  & 6.18 - 8.98 - 16.46  & 40.5  & 5.0  & 39.4 - 13.3 - 5.7 \tabularnewline
{ 5 }  & C/1976 D2  & 6.88  & 44.5  & 5001  & 0  & 0  & 30.2 - 35.1 - 42.2  & 4.76 - 5.50 - 6.00  & 56.9  & 7.3  & 0 - 0 - 0 \tabularnewline
{ 6 }  & C/1976 U1  & 5.86  & 30.5  & 4556$^{a}$  & 445  & 81  & 26.7 - 41.8 - 78.4{[}R{]}  & 5.51 - 5.72 - 56.21{[}R{]}  & 45.6  & 21.9  & 32.2 - 27.1 - 24.3 \tabularnewline
{ 9 }  & C/1979 M3  & 4.69  & -14.2  & 4781$^{a}$  & 220  & 13  & 33.2 - 47.9 - 79.1{[}R{]}  & 4.30 - 4.48 - 24.35{[}R{]}  & 41.1  & 14.5  & 20.9 - 16.9 - 15.0 \tabularnewline
{10 }  & {C/1980 E1$^{{\rm NG}}$}  & 3.36  & 13.3  & 5001  & 0  & 0  & 34.3 - 37.5 - 41.3  & 1.82 - 2.16 - 2.41  & 53.3  & 3.9  & 0 - 0 - 0 \tabularnewline
{11 }  & {C/1983 O1$^{{\rm NG}}$}  & 3.32  & -55.2  & 4503$^{a}$  & 498  & 210  & 18.2 - 30.9 - 68.6  & 3.53 - 4.74 - 52.74  & 60.8  & 36.1  & 32.8 -27.6 - 24.8 \tabularnewline
{13 }  & C/1987 F1  & 3.62  & 26.5  & 5001  & 0  & 0  & 31.0 - 34.4 - 38.5  & 4.82 - 5.40 - 6.39  & 58.2  & 5.0  & 0.1 - 0 - 0 \tabularnewline
{14 }  & C/1987 H1  & 5.46  & 17.6  & 5001  & 0  & 0  & 40.8 - 44.0 - 47.8  & 8.66 - 9.79 - 11.53  & 45.5  & 2.8  & 42.5 - 0.3 - 0 \tabularnewline
{18 }  & C/1993 F1  & 5.90  & 45.1  & 5001  & 0  & 0  & 28.3 - 31.9 - 36.7  & 7.23 - 8.00 - 9.48  & 62.6  & 6.3  & 5.8 - 0.1 - 0 \tabularnewline
{22 }  & {C/1997 J2$^{{\rm NG}}$}  & 3.05  & 0.5  & 5001  & 0  & 0  & 44.8 $\pm$ 0.9  & 2.801 $\pm$ 0.017  & 44.6  & 0.9  & 0 - 0 - 0 \tabularnewline
{24 }  & C/1999 F2  & 4.72  & 46.0  & 5001  & 0  & 0  & 39.1 - 42.6 - 46.8  & 6.21 - 7.03 - 8.59  & 46.9  & 3.3  & 2.3 - 0.1 - 0 \tabularnewline
{25 }  & {C/1999 H3$^{{\rm NG}}$}  & 3.50  & 49.9  & 5001  & 0  & 0  & 16.0 $\pm$ 0.5  & 3.629 $\pm$ 0.014  & 124.7  & 3.9  & 0 - 0 - 0 \tabularnewline
{28 }  & C/1999 N4  & 5.51  & 23.1  & 5001  & 0  & 0  & 29.4 $\pm$ 0.7  & 6.527 $\pm$ 0.092  & 68.2  & 1.7  & 0 - 0 - 0 \tabularnewline
{29 }  & C/1999 S2  & 6.47  & -22.3  & 5001  & 0  & 0  & 32.6 - 35.4 - 38.8  & 8.17 - 8.79 - 9.77  & 56.4  & 3.8  & 6.6 - 0 - 0 \tabularnewline
{30 }  & C/1999 U1  & 4.14  & -25.0  & 5001  & 0  & 0  & 48.0 - 52.8 - 58.8  & 6.23 - 7.84 - 11.39  & 37.9  & 3.0  & 18.6 - 2.5 - 0 \tabularnewline
{32 }  & {C/1999 Y1$^{{\rm NG}}$}  & 3.09  & -62.7  & 5001  & 0  & 0  & 42.2 $\pm$ 0.8  & 5.82 - 6.11 - 6.46  & 47.4  & 0.9  & 0 - 0 - 0 \tabularnewline
{34 }  & {C/2000 CT$_{54}^{{\rm NG}}$}  & 3.16  & -36.7  & 5001  & 0  & 0  & 27.4 $\pm$ 1.2  & 3.88 - 4.04 - 4.24  & 72.9  & 3.1  & 0 - 0 - 0 \tabularnewline
{36 }  & C/2000 O1  & 5.92  & -23.5  & 5001  & 0  & 0  & 34.4 - 38.5 - 43.5  & 7.16 - 7.87 - 9.19  & 52.0  & 4.7  & 4.1 - 0 - 0 \tabularnewline
{37 }  & {C/2000 SV$_{74}^{{\rm NG}}$}  & 3.54  & 21.7  & 5001  & 0  & 0  & 21.7 $\pm$ 0.2  & 3.791 $\pm$ 0.009  & 92.3  & 0.8  & 0 - 0 - 0 \tabularnewline
{45 }  & C/2002 J5  & 5.73  & 33.0  & 5001  & 0  & 0  & 33.77 $\pm$ 0.39  & 7.966 $\pm$ 0.098  & 59.2  & 0.7  & 0 - 0 - 0 \tabularnewline
{47 }  & {C/2002 R3$^{{\rm NG}}$}  & 3.87  & -45.8  & 5001  & 0  & 0  & 38.3 - 41.5 - 45.1  & 4.69 - 5.15 - 5.94  & 48.2  & 3.1  & 0 - 0 - 0 \tabularnewline
{55 }  & {C/2005 EL$_{173}^{{\rm NG}}$}  & 3.89  & -23.2  & 5001  & 0  & 0  & 43.66 $\pm$ 0.99  & 7.82 - 8.31 - 8.89  & 44.8  & 1.0  & 0 - 0 - 0 \tabularnewline
{61 }  & {C/2006 S2$^{{\rm NG}}$}  & 3.16  & -14.9  & 5001  & 0  & 0  & 24.1 - 27.5 - 32.2  & 3.193 - 3.221 - 3.294  & 72.5  & 8.1  & 0 - 0 - 0 \tabularnewline
{63 }  & C/2007 JA$_{21}$  & 5.37  & 31.4  & 5001  & 0  & 0  & 30.75 $\pm$ 0.95  & 5.86 - 5.94 - 6.04  & 65.1  & 2.0  & 0 - 0 - 0 \tabularnewline
\hline
\end{tabular}
\end{table*}

\subsection{Detection of the non-gravitational forces in large perihelion distance
LPCs}

\label{sub:Non-gravitational-forces-detection}

It is well-known that the actual motion of the comet is not purely
gravitational (here and after GR). This means, that very small uncertainties of orbital
elements in a GR model are often only the formal errors
calculated in the process of orbit determination. These small uncertainties,
in turn, may lead us to believe that we know the true orbit of a comet
so perfectly. When incorporating NG effects in the
dynamical model we often obtain a slightly larger formal uncertainties
but we believe that the resulting orbit better represents the real
cometary motion, especially when extrapolating out of the observed
time interval. In particular, this may be the case of some near-parabolic
orbit when taking into account NG~effects may even lead to changes
in the shape of the original barycentric orbit from hyperbolic to
elliptical (\citealp{krolikowska:2001,krolikowska:2006a}, and Fig.~5
in Paper~I). It has been shown that almost all comets with hyperbolic
original GR orbits have original NG~orbits elliptical.

In Paper~I, our sample of comets with the NG~effects was composed
of objects with NG~orbits characterized by a clear decrease in the
rms compared to their rms for purely gravitational orbits. We found
then 26 comets with NG~orbits very well determined for which original
reciprocals of semimajor axes in the NG~model, $1/a_{{\rm ori,NG}}$,
were located within the range from zero to $100\times10^{-6}$\,au$^{-1}$.
In preparing that sample many comets of a purely GR orbit
in Oort spike has been excluded because their NG orbit gave $1/a_{{\rm ori,NG}}>100\times10^{-6}$\,au$^{-1}$.
That sample of 26~comets contained only four objects with perihelion
distances exceeding 3\,au.

In the present study we approach the issue differently. Emphasis is
laid on the completeness of the sample. This allowed us not only to
investigate the dynamic behaviour in the previous and future perihelion
but also to make a statistical analysis of the observed sample of
Oort spike comets and their apparent source region. Knowing that the
NG~effects in the motion of LP~comets are hardly determined we concentrated
on a sample of comets with large perihelion distances in order to
keep the highest possible orbit reliability even for objects with
indeterminable NG~effects.

However, we have not given up on determining the NG~effects in these
comets though, for comets with large perihelion distances the NG~effects
are determined even more difficult than for comets approaching closer
to the Sun. Therefore, we decided to set out the NG~orbit regardless
of whether it is a significant drop in rms. We assumed that even if
this decrease is negligible, the model of the NG~motion should give
a better representation of the actual comet's orbit and its uncertainty
(NG~model consists from one to four additional parameters, so the
uncertainties of orbital elements can be significantly larger). Once
determined, the NG~parameters with reasonable accuracy, we recognized
the model as more realistic than the purely GR model. The
obtained NG~models were also reviewed for O-C~time variations
and O-C distributions (a more detailed description of the methods
used are given in Paper~I). In this way, from the sample of 64~comets,
we have determined NG~orbits for 15~objects with perihelion distances
greater than 3.0\,au. It appeared that all of them have q lower or
equal to 4\,au. Comparison between GR~and NG~solutions for these
fifteen comets is given in Table~\ref{tab:a_original_15} where comets
are ordered by discovery date. Four comets
from Paper~I, with a very well-determined NG~effects are included here. Finally, in
our sample, we have 23~objects with perihelion distance in the range
of 3.0\,au\,$<$\,$q$\,$<$\,4.0\,au (see also Table
\ref{tab:sample_q_dist}). It means that only eight comets of this
perihelion distance range still seem to have undetectable NG~effects.
It should be summarized that it is clearly more than a half
of all comets with perihelion distances between 3 and 4 au (discovered
after 1970) that show small deviations from purely gravitational motion,
which can be detectable either through a decrease in rms for NG model
of motion or by improvements when analysing the differences in OC
distribution and/or OC time variations between GR and NG models.

To determine the NG~cometary orbit we used the same formalism as
in Paper~I that was originally proposed by \citet{marsden-sek-ye:1973}.
This formalism introduces three orbital components of the NG~acceleration
($A_{1},A_{2}$, and $A_{3}$, i.e. radial, transverse and normal
components, respectively) acting on a comet in a case of sublimation
symmetric relative to perihelion. The asymmetric NG~model introduces
additional parameter $\tau$, the time displacement of the maximum
of the g(r$(t-\tau)$). From orbital calculations, the
NG~parameters $A_{1},A_{2}$, and $A_{3}$ and eventually $\tau$
should be derived together with six keplerian orbital elements within
a given observational time interval (more details are given in \citealp{krolikowska:2006a}).
We realize that the standard g(r)-function used by us has been obtained
by \citet{marsden-sek-ye:1973} phenomenologically on the basis of
water sublimation. We have tried to determine the NG~orbits for the
investigated comets assuming more general form of g(r)-like function
(see \citealp{krolikowska:2004}), however, it appeared that the rms
and OC~time variations were practically the same. So we
came to the conclusion that it is best to use standard form of the
g(r)-function. Past experience in determining the NG~orbit on the
basis of various g(r)-like functions (\citealp{krolikowska:2004})
justifies such an approach.

For six comets studied here we were able to derive all three parameters
of NG~acceleration, for the next eight - the radial and the transverse
component of the NG~acceleration, in one case (C/1984~W2) -- only
the radial term.

\subsection{Original and future orbits. Creating swarms of virtual comets.}

\label{sub:Original-and-future}The very first step in studying past
and future motion of LP~comets is to determine their \emph{original}
and \emph{future} orbits. As usual, we call an orbit \emph{original}
when traced back out of reach of planetary perturbations (assumed
in this paper to happen at 250\,au from the Sun). Similarly, the
\emph{future} orbit can be obtained after following the motion of
a comet up in time as far as it reaches the same heliocentric distance
of 250\,au.

To derive original and future orbital parameters (including semimajor
axes) as well as their uncertainties we have examined the evolution
of thousands of virtual comet (VC) orbits using the Sitarski's method
of the random orbit selection \citep{sitarski:1998}. More details
are given in Paper~I. With this method, we construct osculating swarms
of comets that follow the normal distribution in the space of orbital
elements and eventually NG~parameters (the 6--9~dimensional normal
statistics). Similarly as in Paper~I we fill a confidence region
with 5\,000~VCs for each nominal solution (GR and NG if determined)
for each considered comet.

Next, each of VC from the osculating swarm was followed numerically
from its position at osculation epoch backwards and forwards until
this individual VC reached a distance of 250\,au from the Sun. The
equations of comet's motion have been integrated numerically using
the recurrent power series method \citep{sitarski:1989,sitarski:2002},
taking into account perturbations by all the planets and including
the relativistic effects. In this way we were able to obtain the nominal
original and future barycentric orbits of each comet as well as the
uncertainties of the derived values of orbital elements by fitting
the normal distribution to each original and future cometary swarm
\citep{krolikowska:2001}. Obtained original and future reciprocals
of semimajor axes with their uncertainties are given in Tables~\ref{tab:past_motion_old}
-- \ref{tab:past_motion_uncertain} and Tables~\ref{tab:future_motion_returning}--\ref{tab:future_motion_escaping}
where comets are ordered by discovery date. As it was mentioned before,
only one comet in the investigated sample, C/1978~G2, formally have
a negative $1/a_{{\rm ori}}$ but large uncertainty of this
value clearly does not exclude Solar System origin of this comet.

\vspace{0.2cm}

The original and future swarms of VCs became the starting data to
study the dynamical evolution of each individual comet under the Galactic
tides (Section~\ref{sec:Prev-and-next-perihelion-passages}).

\begin{figure}
\begin{centering}
\includegraphics[width=8.6cm]{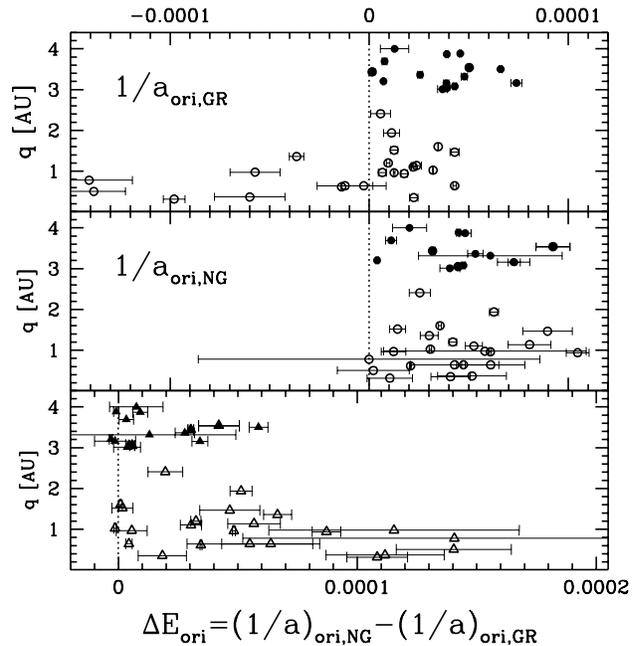} 
\par\end{centering}

\caption{Shifts of $1/a_{{\rm ori}}$ due to the NG~effects for 15 of
investigated comets and all comets from Paper~I.}

\centering{}\label{fig:spik_ng} 
\end{figure}

\subsection{Differences between non-gravitational and gravitational models}

\label{sub:NG-and-GR-models}

For 15 comets with the determined NG~orbits we present only the results
for the dynamic evolution of NG~swarms of VCs (Tables~\ref{tab:past_motion_old}
-- \ref{tab:past_motion_uncertain} and Tables~\ref{tab:future_motion_returning}--\ref{tab:future_motion_escaping}).
However, we derived also their original and future orbits starting
from the GR osculating orbits. Differences between the
results obtained we will discuss in terms of differences in values
of $1/a_{{\rm ori}}$ and $1/a_{{\rm fut}}$, and possible differences
in the assessment of the dynamical status of investigated comets (Section~\ref{sec:New-and-old}).
In this section we will focus on the first issue.

Figure~\ref{fig:spik_ng} shows the differences between original
reciprocals of semimajor axes for GR and
NG~models. Comets investigated here are presented by filled symbols
(15 comets with $q\geq3.0$\,au), while the open symbols represent
22~comets with $q<3.0$\,au from Paper~I. When drawing conclusions,
remember that comet sample analysed in the Paper~I was constructed
on the other principles than a sample of comets currently being examined.
However, Fig.~\ref{fig:spik_ng} clearly shows that in the case of
comets with low $q$ (less than, say, $q\leq2$\,au), the differences
$\Delta(1/a_{{\rm ori}})=1/a_{{\rm ori,NG}}-1/a_{{\rm ori,GR}}$
are much greater than that for comets with large $q$ ($q$ greater than
3\,au). For more than 50 per cent of large perihelion comets $\Delta(1/a_{{\rm ori}})<20\times10^{-6}$\,au
$^{-1}$. It is worth noting that of 15~comets with $1/a_{{\rm ori,GR}}$
within Oort spike, only one has an NG~orbit slightly outside the
peak (C/1999~H3, $1/a_{{\rm ori,NG}}=(124.66\pm3.88)\times10^{-6}$\,au\,$^{-1}$)
while having the greatest value of $\Delta(1/a_{{\rm ori}})=58\times10^{-6}$\,au\,$^{-1}$
resulting from planetary perturbations. When constructing the sample
of NG~Oort spike comets in Paper~I, we obtained several such cases.

Usually the uncertainties of orbital elements including original and
future semimajor axes are larger for NG~orbits than GR~orbits. Thus,
initial NG~swarms of VCs for original and future orbit calculations
are more dispersed than GR~swarms. In our opinion, the NG~swarms
of VCs better reflect actual knowledge of studied cometary orbits.
This affects the determination of $1/a_{{\rm ori,NG}}$ uncertainty,
which is generally a factor of 2--3 larger than the uncertainty of
$1/a_{{\rm ori,GR}}$ determination for the same comet, as it
can be observed in Table \ref{tab:a_original_15}. An extreme case
is $1/a_{{\rm ori,NG}}=(60.8\pm36.1)\times10^{-6}$\,au$^{-1}$
for comet C/1983~O1 while the purely GR~swarm gives $1/a_{{\rm ori,GR}}=(47.8\pm1.6)\times10^{-6}$\,au$^{-1}$,
which is solution formally more than one order of magnitude more accurate.

For the remaining 49 comets NG~effects are indeterminable. Hence,
we present the results for the purely gravitational swarms of VCs.
However, only 8 of these comets have perihelion distances less than
4\,au. With the apparent trend of significant reduction of differences
between NG~orbit and GR~orbit with the increasing perihelion distance
(Fig.~\ref{fig:spik_ng}) it seems reasonable that the vast majority
of real orbits of these 49~comets is well-targeted despite of omitting
the NG~effects.

\begin{figure}
\includegraphics[height=1\columnwidth,angle=-90]{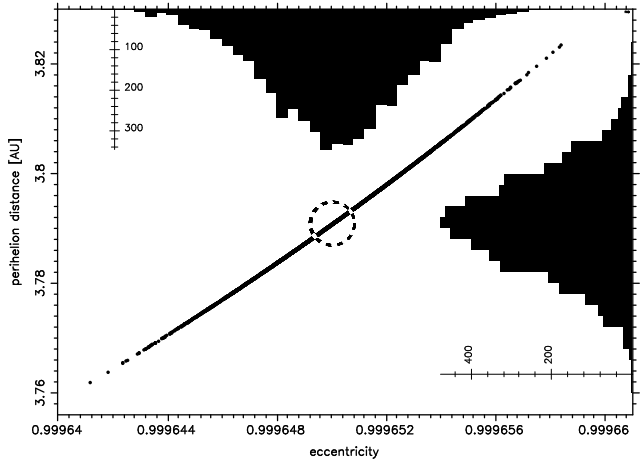}
\caption{\label{fig:Past-distrib_2000sv}Previous \emph{perihelion distance
- eccentricity} distribution for C/2000~SV$_{74}$. All 5001 VCs
were stopped at the previous perihelion. The centre of the dashed
circle marks the nominal orbit.}

\end{figure}

\begin{figure}
\includegraphics[height=1\columnwidth,angle=-90]{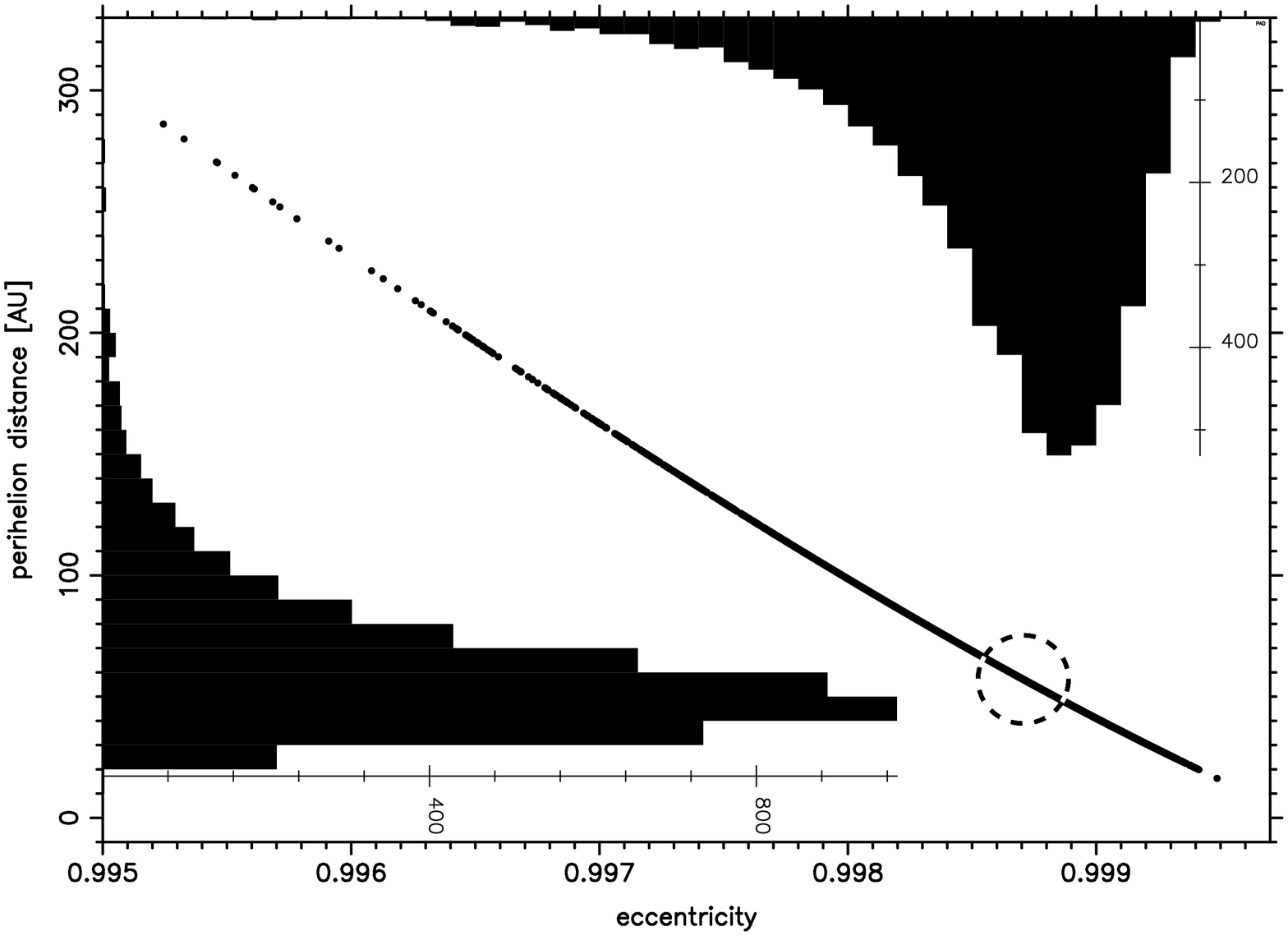}
\caption{\label{fig:Past-distrib_2005q1}Previous \emph{perihelion distance
- eccentricity} distribution for C/2005~Q1. All 5001~VCs were stopped
at the previous perihelion. The centre of the dashed circle marks
the nominal orbit.}

\end{figure}

\begin{figure}
\includegraphics[bb=14bp 14bp 471bp 346bp,width=1\columnwidth]{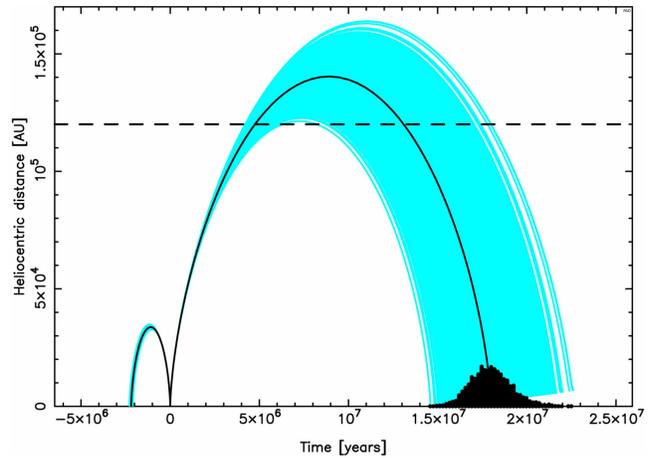}

\caption{\label{fig:fon_2002j5}Past and future heliocentric distance changes
of all VCs representing C/2002~J5. Grey (cyan) lines present heliocentric
distance changes of all VCs except the nominal ones, for which black
line is used. Additionally the distribution of the future VCs swarm
(black histogram) in time is presented.}

\end{figure}

\begin{table*}
\caption{\label{tab:past_motion_new}The past motion of \emph{dynamically new}
comets. The table is organized in the same manner as Table \ref{tab:past_motion_old}.
In the case of mixed swarm the mean values or deciles of $Q$ and
$q$ are given for the returning part of the VCs swarm, where the
escape limit of 120\,000\,au was generally adopted with the exception
of four objects mark with an asterisk (C/2001~C1 and C/2004~X3)
or two asterisks (C/2001~K5 and C/2003~G1), where the escape limit
of 140\,000\,au and 200\,000\,au was applied, respectively.}

\centering{}\begin{tabular}{rlcccccccr@{$\pm$}rc}
\hline 
\#  & Comet  & $q_{{\rm osc}}$  & $b_{{\rm osc}}$  & \multicolumn{3}{c}{Number of VCs} & $Q_{{\rm prev}}$  & $q_{{\rm prev}}$  & \multicolumn{2}{c}{$1/a_{{\rm ori}}$} & \% of \tabularnewline
 &  & au  & deg  & {[}R{]}  & {[}E{]}  & {[}H{]}  & $10^{3}$au  & au  & \multicolumn{2}{c}{$10^{-6}$au$^{-1}$} & dyn. new \tabularnewline
 &  &  &  &  &  &  &  &  & \multicolumn{2}{c}{ } & 10\,au - 15\,au - 20\,au \tabularnewline
{[}1{]}  & {[}2{]}  & {[}3{]}  & {[}4{]}  & {[}5{]}  & {[}6{]}  & {[}7{]}  & {[}8{]}  & {[}9{]}  & \multicolumn{2}{c}{{[}10{]}} & {[}11{]} \tabularnewline
\hline 
{ 4 }  & {C/1974 V1}  & 6.02  & -28.0  & 2739$^{a}$  & 2262  & 337  & 54.1 - 80.6 - 112.7{[}R{]}  & 12.49 - 78.23 -576.36{[}R{]}  & 17.7  & 12.1  & \textbf{96.9 - 92.0 - 88.8} \tabularnewline
{ 7 }  & {C/1978 A1}  & 5.61  & 31.4  & 4814$^{a}$  & 187  & 5  & 37.8 - 52.7 - 82.2{[}R{]}  & 9.01 - 19.52 - 141.32{[}R{]}  & 37.5  & 11.9  & \textbf{85.1 - 63.3 - 50.9} \tabularnewline
{ 8 }  & {C/1978 G2}  & 6.28  & 34.2  & 792  & 4209  & 3585$^{a}$  & 33.3 - 62.5 - 109.0{[}R{]}  & 7.25 - 27.05 - 568.1{[}R{]}  & -22.4  & 37.8  & \textbf{95.7 - 94.0 - 93.1} \tabularnewline
{12 }  & {C/1984 W2$^{{\rm NG}}$}  & 4.00  & -33.0  & 3488$^{a}$  & 1513  & 43  & 60.8 - 84.4 - 113.1{[}R{]}  & 12.8 - 91.3 - 606.0{[}R{]}  & 20.3  & 8.6  & \textbf{95.1 - 91.6 - 88.0} \tabularnewline
{15 }  & {C/1987 W3}  & 3.33  & -64.7  & 4534$^{a}$  & 467  & 1  & 58.5 - 78.0 - 110.7{[}R{]}  & 14.1 - 61.1 - 451.8{[}R{]}  & 24.7  & 7.3  & \textbf{95.8 - 89.8 - 82.7} \tabularnewline
{16 }  & {C/1988 B1}  & 5.03  & 47.7  & 3770$^{a}$  & 1231  & 7  & 66.7 - 89.6 - 116.9{[}R{]}  & 39.9 - 208.3 - 985.0{[}R{]}  & 20.2  & 7.0  & \textbf{99.9 - 99.1 - 98.1} \tabularnewline
{17 }  & {C/1992 J1}  & 3.00  & 43.7  & 5001  & 0  & 0  & 73.6 $\pm$ 2.5  & 20.4 - 29.7 - 43.9  & 27.2  & 0.9  & \textbf{100 - 99.0 - 91.1} \tabularnewline
{20 }  & {C/1997 A1}  & 3.16  & 19.9  & 4993  & 8  & 0  & 83.4 - 91.6 - 101.2{[}R{]}  & 56.7 - 111.2 - 224.6 {[}R{]}  & 21.8  & 1.7  & \textbf{100 - 100 - 100} \tabularnewline
{21 }  & {C/1997 BA$_{6}^{{\rm NG}}$}  & 3.44  & -27.7  & 5001  & 0  & 0  & 60.1 - 62.8 - 65.8  & 15.9 - 19.5 - 24.7  & 31.8  & 1.2  & \textbf{100 - 95.3} - 44.6 \tabularnewline
{26 }  & {C/1999 J2}  & 7.11  & 49.8  & 5001  & 0  & 0  & 90.4 $\pm$ 2.6  & 160 - 199 - 253  & 22.1  & 0.6  & \textbf{100 - 100 - 100} \tabularnewline
{27 }  & {C/1999 K5}  & 3.26  & -34.0  & 5001  & 0  & 0  & 93.6 $\pm$ 4.1  & 158 - 219 - 314  & 21.4  & 0.9  & \textbf{100 - 100 - 100} \tabularnewline
{31 }  & {C/1999 U4}  & 4.92  & 30.7  & 5001  & 0  & 0  & 62.9 $\pm$ 1.0  & 37.3 $\pm$ 2.7  & 31.8  & 0.5  & \textbf{100 - 100 - 100} \tabularnewline
{33 }  & {C/2000 A1}  & 9.74  & 33.2  & 5001  & 0  & 0  & 49.3 $\pm$ 2.4  & 21.4 - 24.9 - 29.7  & 40.6  & 2.0  & \textbf{100 - 100 - 97.8} \tabularnewline
{39 }  & {C/2001 C1$^{*}$}  & 5.10  & 8.3  & 4019$^{a}$  & 982  & 0  & 106.2 - 122.2 - 136.9{[}R{]}  & 68.9 - 179.0 - 366.0{[}R{]}  & 15.9  & 2.1  & \textbf{100 - 100 - 100} \tabularnewline
{41 }  & {C/2001 K3}  & 3.06  & -17.6  & 4919$^{a}$  & 82  & 0  & 50.2 - 64.0 - 87.0{[}R{]}  & 6.08 - 14.9 - 83.7{[}R{]}  & 31.1  & 6.8  & \textbf{67.8 - 50.6} - 40.2 \tabularnewline
{42 }  & {C/2001 K5}$^{**}$  & 5.18  & 30.2  & 4741$^{a}$  & 260  & 0  & 182.0 $\pm$ 4.8  & 19200 $\pm$ 2300  & 9.6  & 0.4  & \textbf{100 - 100 - 100} \tabularnewline
{43 }  & {C/2002 A3}  & 5.15  & 9.0  & 4942$^{a}$  & 59  & 0  & 86.3 - 95.9 - 107.6  & 48.5 - 82.9 - 156.2  & 20.7  & 1.8  & \textbf{100 - 100 - 100} \tabularnewline
{44 }  & {C/2002 J4}  & 3.63  & -32.5  & 5001  & 0  & 0  & 58.8 $\pm$ 2.4  & 22.5 - 27.7 - 35.3  & 34.1  & 1.4  & \textbf{100 - 100 - 97.7} \tabularnewline
{46 }  & {C/2002 L9}  & 7.03  & -51.7  & 5001  & 0  & 0  & 54.7 $\pm$ 1.3  & 19.4 - 21.4 - 24.1  & 36.5  & 0.9  & \textbf{100 - 100 - 80.9} \tabularnewline
{48 }  & {C/2003 G1}$^{**}$  & 4.92  & 21.9  & 5001  & 0  & 0  & 136.6 - 144.4 - 152.9  & 2041 - 2870 - 4056  & 13.7  & 0.6  & \textbf{100 - 100 - 100} \tabularnewline
{49 }  & {C/2003 S3}  & 8.13  & -11.5  & 5001  & 0  & 0  & 50.3 - 55.5 - 62.1  & 14.0 - 16.8 - 22.1  & 36.0  & 2.9  & \textbf{100 - 78.0} - 19.6 \tabularnewline
{51 }  & {C/2004 P1}  & 6.01  & -20.2  & 5001  & 0  & 0  & 57.0 - 64.1 - 73.3  & 18.1 - 27.8 - 51.0  & 31.1  & 3.0  & \textbf{100 - 97.9 - 82.4} \tabularnewline
{52 }  & {C/2004 T3}  & 8.86  & -34.7  & 5001  & 0  & 0  & 39.1 - 43.8 - 49.8  & 13.4 - 15.9 - 20.7  & 45.7  & 4.2  & \textbf{100 - 64.8} - 12.5 \tabularnewline
{53 }  & {C/2004 X3$^{*}$}  & 4.40  & 36.6  & 2772$^{a}$  & 2229  & 0  & 118.6 - 134.0 - 144.9{[}R{]}  & 746 - 1677 - 2755{[}R{]}  & 13.4  & 2.1  & \textbf{100 - 100 - 100} \tabularnewline
{54 }  & {C/2005 B1$^{{\rm NG}}$}  & 3.20  & 17.8  & 0  & 5001  & 0  & --  & --  & 4.0  & 0.6  & \textbf{100 - 100 - 100} \tabularnewline
{56 }  & {C/2005 G1}  & 4.96  & 55.3  & 4551$^{a}$  & 450  & 0  & 119.0 $\pm$ 6.0{[}R{]}  & 490 - 739 - 1046{[}R{]}  & 16.6  & 1.0  & \textbf{100 - 100 - 100} \tabularnewline
{57 }  & {C/2005 K1$^{{\rm NG}}$}  & 3.69  & 17.1  & 112  & 4889$^{a}$  & 0  & 106.5 - 114.9 - 118.8{[}R{]}  & 207.2 - 365.1 - 465.2{[}R{]}  & 11.0  & 2.7  & \textbf{100 - 100 - 100} \tabularnewline
{58 }  & {C/2005 Q1}  & 6.41  & 8.9  & 5001  & 0  & 0  & 78.8 - 87.8 - 99.7  & 34.6 - 56.0 - 105.3  & 22.7  & 2.0  & \textbf{100 - 100 - 100} \tabularnewline
{59 }  & {C/2006 E1}  & 6.04  & -31.5  & 5001  & 0  & 0  & 56.6 - 61.5 - 67.6  & 21.1 - 29.3 - 45.0  & 32.5  & 2.2  & \textbf{100 - 99.8 - 93.3} \tabularnewline
{60 }  & {C/2006 K1}  & 4.43  & -64.0  & 4621$^{a}$  & 376  & 0  & 114.0 - 121.7 - 129.2{[}R{]}  & 375 - 558 - 793{[}R{]}  & 16.2  & 0.9  & \textbf{100 - 100 - 100} \tabularnewline
{64 }  & {C/2007 Y1}  & 3.34  & 35.6  & 4631$^{a}$  & 370  & 10  & 39.2 - 56.8 - 90.0{[}R{]}  & 4.6 - 14.5 - 230.0{[}R{]}  & 34.1  & 12.5  & \textbf{63.3 - 52.9} - 47.3 \tabularnewline
\hline
\end{tabular}
\end{table*}

\begin{table*}
\caption{\label{tab:past_motion_uncertain}The past motion of 7 comets with
uncertain \emph{new/old} status. The table is organized in the same
manner as Table \ref{tab:past_motion_old}. In the case of mixed swarm
the mean values or deciles of $Q$ and $q$ are given for the returning
part of the VCs swarm. The escape limit of 120\,000\,au was used
for all these comets.}

\begin{tabular}{rlcccccccr@{$\pm$}rc}
\hline 
\#  & Comet  & $q_{{\rm osc}}$  & $b_{{\rm osc}}$  & \multicolumn{3}{c}{Number of VCs} & $Q_{{\rm prev}}$  & $q_{{\rm prev}}$  & \multicolumn{2}{c}{$1/a_{{\rm ori}}$} & \% of \tabularnewline
 &  & au  & deg  & {[}R{]}  & {[}E{]}  & {[}H{]}  & $10^{3}$au  & au  & \multicolumn{2}{c}{$10^{-6}$au$^{-1}$} & dyn. new \tabularnewline
 &  &  &  &  &  &  &  &  & \multicolumn{2}{c}{ } & 10\,au - 15\,au - 20\,au \tabularnewline
{[}1{]}  & {[}2{]}  & {[}3{]}  & {[}4{]}  & {[}5{]}  & {[}6{]}  & {[}7{]}  & {[}8{]}  & {[}9{]}  & \multicolumn{2}{c}{{[}10{]}} & {[}11{]} \tabularnewline
\hline 
{19 }  & {C/1993 K1}  & 4.85  & -2.3  & 4753$^{a}$  & 248  & 2  & 51.2 - 68.3 - 94.6{[}R{]}  & 6.33 - 10.14 - 32.89{[}R{]}  & 29.1  & 7.7  & \textbf{53.4} - 33.0 - 23.8 \tabularnewline
{23 }  & {C/1999 F1}  & 5.79  & -15.2  & 5001  & 0  & 0  & 53.5 $\pm$ 1.0  & 12.12 $\pm$ 0.48  & 37.4  & 0.7  & \textbf{100} - 0 - 0 \tabularnewline
{35 }  & {C/2000 K1}  & 6.28  & 21.5  & 5001  & 0  & 0  & 50.2 $\pm$ 2.9  & 12.6 - 15.0 - 18.5  & 40.0  & 2.3  & \textbf{100} - 49.3 - 4.8 \tabularnewline
{38 }  & {C/2000 Y1}  & 7.97  & -28.0  & 5001  & 0  & 0  & 30.4 - 33.1 - 36.4  & 9.8 - 10.5 - 11.6  & 60.4  & 4.2  & \textbf{80.9} - 0 - 0 \tabularnewline
{40 }  & {C/2001 G1}  & 8.24  & 46.8  & 5001  & 0  & 0  & 49.5 $\pm$ 3.4  & 12.4 - 14.5 - 18.0  & 40.6  & 2.8  & \textbf{100} - 40.7 - 4.4 \tabularnewline
{50 }  & {C/2003 WT$_{42}$}  & 5.19  & 57.0  & 5001  & 0  & 0  & 44.03 $\pm$ 0.31  & 12.45 $\pm$ 0.22  & 45.4  & 0.3  & \textbf{100} - 0 - 0 \tabularnewline
{62 }  & {C/2006 YC}  & 4.95  & 27.7  & 4965$^{a}$  & 36  & 0  & 32.8 - 43.9 - 64.6{[}R{]}  & 6.5 - 9.9 - 34.6{[}R{]}  & 45.7  & 12.0  & \textbf{49.9} - 28.7 - 20.0 \tabularnewline
\hline
\end{tabular}
\end{table*}

\section{Previous and next perihelion passages}

\label{sec:Prev-and-next-perihelion-passages}

Starting from \emph{original} or \emph{future} LPCs orbits we followed
their dynamical evolution under the Galactic tides, where both disk
and central terms were included. This is possible only in the absence
of any other perturbing forces. \citet{dyb-hab3:2006} have shown,
that none of the known star influenced the motion of LP comets significantly
in the last, say 10~million years, and the same holds for the same
interval in the future. The similar conclusion one can find in \citet{delsemme:1987}, \citet{matese-w:1992}, \citet{wiegert-tre:1999}, \citet{matese-lissauer:2004}, \citet{dyb-asymm:2002}, \citet{emelyanenko_et_al:2007} or recently \citet{kaib-quinn:2009}.
This time interval is comparable with the orbital period of a comet
having the semimajor axis of 50\,000~au therefore we decided to
follow the motion of each comet for one orbital period to the past
and future. Additionally, longer numerical integrations would show
rather artificial cometary motion in cases of previous/next perihelion
in planetary region since the planetary perturbations cannot be taken
into account for obvious reasons. More discussion on the influence
of omitting stellar perturbations on our results one can found in
Section~\ref{sec:Conclusions}. In the way described above. we obtain
previous and next perihelion passage distances or detect past or future
escape, see Table~\ref{tab:overal}. The final orbits obtained from
this calculations we call \emph{previous} and \emph{next}.

\begin{figure}
\includegraphics[bb=14bp 14bp 471bp 346bp,width=1\columnwidth]{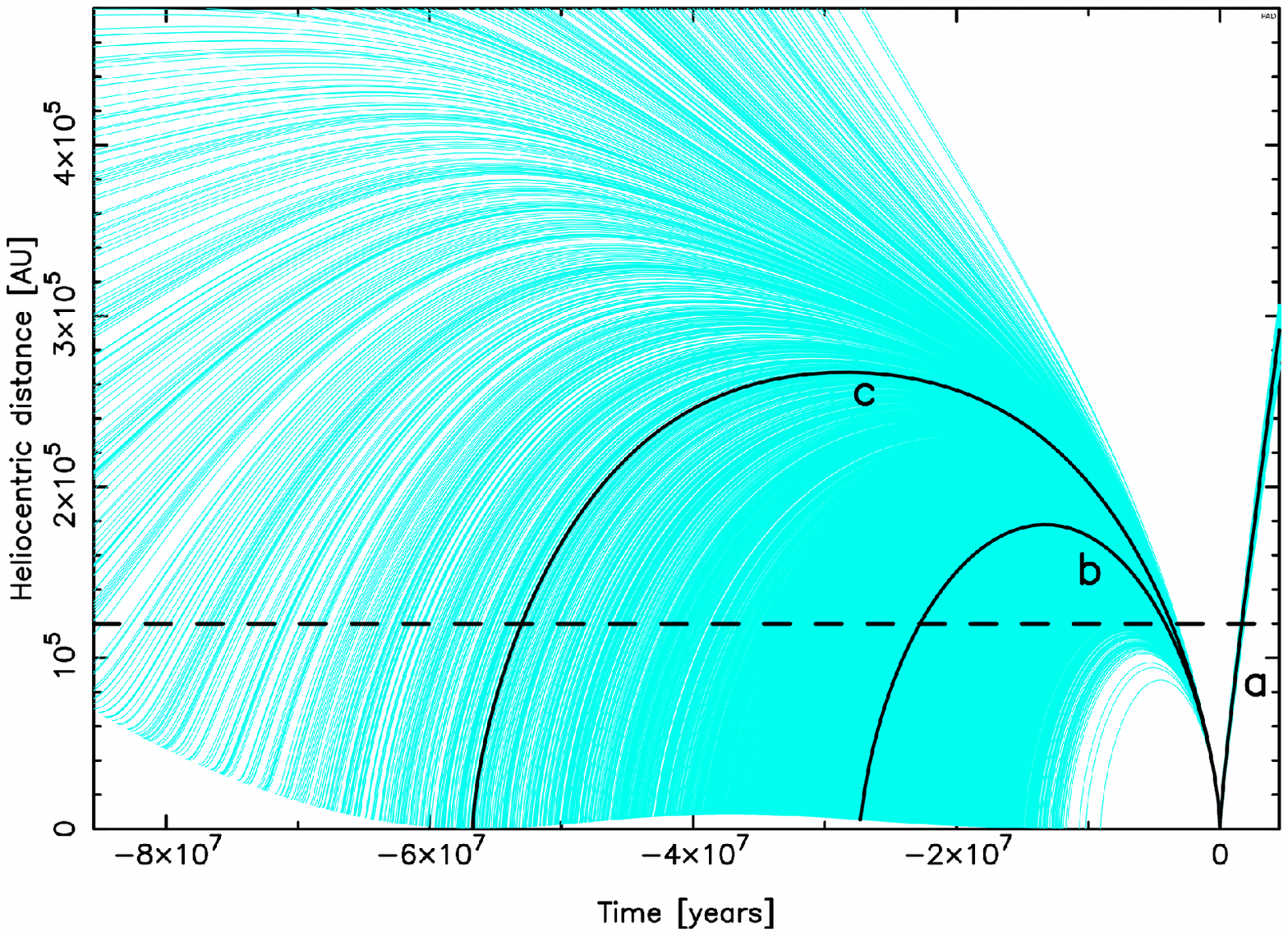}

\caption{\label{fig:fon_2005k1x}Past and future heliocentric distance changes
of all VCs representing C/2005 K1.}

\end{figure}

In this paper we call a comet (or more precisely each individual VC)
as returning {[}R{]} if it goes no further than 120\,000~au from
the Sun. All other comets (or VCs) are called escaping {[}E{]}, but
among them we count escapes in hyperbolic {[}H{]} orbits. For two
comets in our sample, namely C/2001~C1 and C/2004~X3, we decided
to increase the threshold value up to 140\,000\,au because the median
of aphelia of their orbits were slightly below this value (both are
marked with ({*}) in Table \ref{tab:past_motion_new}). For the past
motion of next two comets, C/2001~K5 and C/2003~G1, we present the
results for the escape limit of 200\,000\,au, see Table~\ref{tab:past_motion_new}
where they are marked with ({*}{*}). For such a huge threshold value
all VCs of these comets are returning and we are able to present more
reliable description of the previous perihelion distance then for
the standard threshold, where we are restricted to the synchronous
variant (see below) because all VCs are escaping. The classification
of all these four comets as dynamically new is by no means influenced
by these escape border extensions due to very large previous perihelion
distances of all of them.

For the detail description of the dynamical model as well as its numerical
treatment the reader is kindly directed to Paper~I. Basing on the
conclusions from that work we used both, Galactic disk and Galactic
centre terms in all calculations. Comparing with Paper I all the parameters
of the Galactic gravity field are kept unchanged, including the local
disk mass density, $\rho=0.100$ M$_{\odot}/pc^{3}$.The rules of
stopping the numerical integration were as follows: if all VCs for
a particular comet were returning, all of them were stopped at individual
previous/next perihelia. There were also a synchronous variant, in
which all VCs were stopped simultaneously when the nominal VC reached
previous/next perihelion. When all VCs were escaping the calculation
was always terminated synchronously, when the fastest VC crossed the
escape limit, usually equal to 120\,000\,au. If for a particular
comet the swarm of VCs consists of both returning and escaping VCs,
the returning part was stopped at previous/next VC perihelia and the
rest (escaping ones) when the fastest escaping VC crossed the escape
limit. For these mixed swarms we also performed a synchronous variant,
in which all VCs (both returning and escaping) were halted when the
fastest VC crossed the escape limit.

In Tables~\ref{tab:past_motion_old},\ref{tab:past_motion_new} and
\ref{tab:past_motion_uncertain}, in columns {[}8{]} and {[}9{]} we
presented previous aphelion and perihelion distances, respectively.
In Table~\ref{tab:future_motion_returning} the analogous data for
the comets returning in the future are given in columns {[}7{]} and
{[}8{]}. Depending on the distribution characteristics we use two
different ways of aphelion and perihelion distance presentation: if
the distribution can be reliably approximated with the Gaussian one,
we present the estimated mean value and its standard deviation. The
example of Gaussian distribution of past elements one can find in
Fig.~\ref{fig:Past-distrib_2000sv}. In the case of highly deformed
distribution, we present three deciles: 10th, the median and 90th.
The example of non-Gaussian distribution of past elements one can
find in Fig.~\ref{fig:Past-distrib_2005q1}.

\subsection{Overall statistics}

\label{sub:Overal-statistics}

For the past motion we obtained 42 comets with all VCs returning,
19 comets with the mixed VC~swarms and only 3 comets fully escaping
(but with all VCs on highly eccentric elliptical orbits). For statistics
presented in this section the standard escape limit of 120\,000\,au
for all investigated comets were adopted. Almost all 19~comets with
mixed VC~swarms have the majority of returning clones, only two swarms
consist mainly of escaping VCs (C/2005~K1 and C/1978~G2, the last
comet is the only one with nominal hyperbolic original orbit).

In total, for the past motion of studied comets we obtained the 275\,042 returning VCs
(87.3 per cent) in 315\,063 of all starting from the Oort spike (outside
Oort spike was NG~swarm of C/1999~H3). Statistics of previous perihelion
and aphelion distributions for all these returning VCs are shown in
Table~\ref{tab:overal}.

\begin{table}
\caption{\label{tab:overal}Overall VC distributions for previous and next
perihelion passage (based on the standard escape limit of 120\,000\,au)
for all investigated comets except C/1999~H3, see text.}

\centering{}\begin{tabular}{cccc}
\hline 
deciles  & 10 per cent  & 50 per cent  & 90 per cent \tabularnewline
\hline 
$q_{{\rm previous}}$  & 4.11\,au  & 11.83\,au  & 129.60\,au \tabularnewline
$Q_{{\rm previous}}$  & 29\,000\,au  & 48\,400\,au  & 92\,400\,au \tabularnewline
time to previous perihelion  & 9.66 Myr  & 3.74 Myr  & 1.82 Myr\tabularnewline
\hline 
$q_{{\rm next}}$  & 3.00\,au  & 4.85\,au  & 8.30\,au \tabularnewline
$Q_{{\rm next}}$  & 3\,370\,au  & 7\,080\,au  & 42\,000\,au \tabularnewline
time to next perihelion  & 0.076 Myr  & 0.217 Myr  & 3.027 Myr\tabularnewline
\hline
\end{tabular}
\end{table}

It should be stressed however, that we call escaping all VCs moving
further than 120\,000~au from the Sun. This is motivated by the
fact, that because of a large heliocentric distance and huge
orbital period these VCs may have their orbits modified by (even weak)
stellar perturbations in the past (or future). The only possibility
is to state, that at the moment their dynamical history is impossible
to be revealed. But one should not interpret these comets as of interstellar
origin. The vast majority of the VCs escaping in the past still move
in elliptical, heliocentric orbits and we have no direct evidence
that they were not the Solar System members. See Section \ref{sub:Extremely-large-past-semimajor-axis}
for a detailed analysis of some particular examples.

It is widely known, that the situation is quite different as it concerns
the future motion. In Paper I the great majority of the investigated
22 comets with $q<3.0$\,au (about 77 per cent) were ejected from
the Solar System by planetary perturbations. In the present sample
of 64 comets this percentage is significantly smaller: 33 comets (about
52 per cent) are ejected in the future, two comets from Table~\ref{tab:future_motion_returning}
(see below) and all from Table~\ref{tab:future_motion_escaping}.

We obtained 31 comets with all VCs escaping in the future on hyperbolic
orbits (with the exception of a small part of the VC swarm of C/1987~H1
escaping on the extremely eccentric elliptical orbits). We also obtained
27 comets fully returning in the future and only 6 comet with the
mixed swarms. All these mixed swarms mainly consist of escaping VCs,
nominal VC of two comets have hyperbolic future orbit. Thus, it seems
probable that these two comets (C/1978~G2 and C/2006~S2, see Table~\ref{tab:future_motion_returning})
also are escaping from the solar system. In case of two comets with
the mixed future swarm (but without any hyperbolic VCs, namely C/1997~J2
and C/2002~J5) it is possible to obtain a fully returning future
swarm by applying the escape threshold enlarged up to 200\,000~au.
For C/2002~J5 this is illustrated in Fig.~\ref{fig:fon_2002j5},
where we present the heliocentric distance changes of all VCs representing
C/2002~J5, both one orbital period to the past and future. In contrast
to Fig.\ref{fig:fon_2005k1x} the past VCs swarm is very tight here
and the future swarm, while crossing the standard escape border of
120\,000~au, is all returning, having the future perihelion distance
greater than 2\,500\,au.

For the future motion, we have in total 135\,661 (43.1 per cent)
returning VCs and their statistics is presented in lower part of Table~\ref{tab:overal}.

\begin{figure}
\includegraphics[height=1\columnwidth,angle=-90]{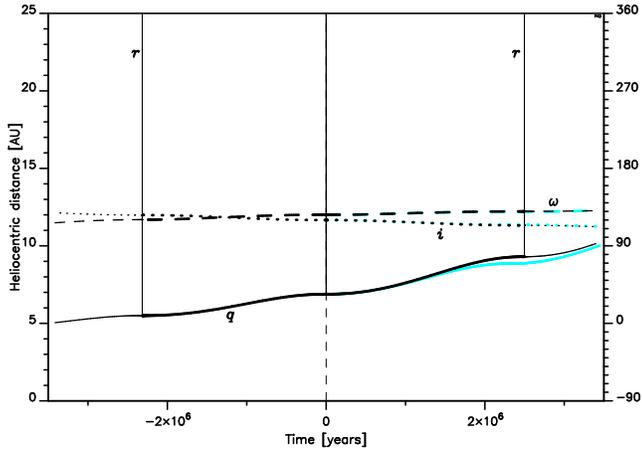}

\caption{\label{fig:gal_1976d2}Past and future orbital evolution of the nominal
VC for C/1976~D2, an example of Oort spike comet detectable at three
consecutive perihelion passages. }

\end{figure}

\subsection{Comets with extremely large semimajor axes in the past}

\label{sub:Extremely-large-past-semimajor-axis}

In our sample, we have six comets with extremely large original semimajor
axes ($1/a_{{\rm ori}}<15\times10^{-6}$\,au$^{-1}$). Three of them,
C/2001~K5, C/2003~G1 and C/2005~B1, have well-determined $1/a_{{\rm ori}}$
and completely escaping swarms of VCs for the standard escape limit
of 120\,000\,au used in this paper. The first two have similar osculating
perihelion distance of about 5\,au and a very similar behaviour in
the past. It can be seen from Table~\ref{tab:past_motion_new} that
a swarm of C/2003~G1 is completely returning for the escape limit
shifted to 200\,000\,au with the previous perihelion of about a
few thousand au from the Sun. The same is true in the case of comet
C/2001~K5 for slightly larger escape border but in this case the
previous perihelion distance is almost one order greater. This rather
suggest, that both are Solar System members, but due to their large
semimajor axes and long orbital period we cannot exclude that their
past motion was disturbed by stellar perturbations and as a result
their dynamical history was quite different.

Third comet with completely escaping past swarm, C/2005~B1, has formally
semimajor axis of about $250\pm50$\,thousand~au! This comet is
also unique in the sense that NG~effects determined for this comet
caused elongation rather than shortening of its NG~original semimajor
axis relative to GR~solution (see Section~\ref{sub:Non-gravitational-forces-detection}
and Fig.~\ref{fig:spik_ng}), however the difference between NG and
GR~models is small: $\Delta(1/a)=(1/a)_{{\rm ori,NG}}-(1/a)_{{\rm ori,GR}}=(-3.2\pm0.9)\times10^{-6}$\,au$^{-1}$.
Taking all this into account comet C/2005~B1 seems to be unique and
we cannot even rule out its interstellar origin. It seems worth to
mention that this comet is the only one of six comets considered in
this section that is returning in the future.

\begin{center}

\begin{table*}
\caption{\label{tab:q_increase} Comets with previous perihelion distance smaller
than the observed one. This means that they were observed after the
minimum point in the Galactic evolution of the perihelion distance,
see also Fig.\ref{fig:gal_deltaq}. In column 5 the statistics of
perihelion distance changes of all VCs during the last orbital revolution
is presented by three deciles, and the percentage of negative $\Delta q$
in column 6. Since changes in $q$ are highly correlated with the
Galactic argument of perihelion $\omega$, its evolution is also shown
in columns 7 and 8, as well as the latitude of the perihelion direction
$b_{{\rm ori}}$. For comparison the previous perihelion distance
in the galactic tide disk model only is given in parentheses in the
third column}

{ \begin{tabular}{rcccccrrrr}
\hline 
\#  & { Name }  & $q_{{\rm prev}}$  & $q_{{\rm ori}}$  & $\Delta q$ {[}au{]}  & \% of  & \multicolumn{3}{c}{galactic coordinates}   & future \tabularnewline
 &  & {[}au{]}  & {[}au{]}  & 10\% \quad{}50\% \quad{}90\%  & $\Delta q$<0  & $\omega_{{\rm prev}}$  & $\omega_{{\rm ori}}$  & $b_{{\rm ori}}$  & \tabularnewline
\hline 
{ 1 }  & { 2 }  & { 3 }  & { 4 }  & {5 }  & { 6 }  & { 7 }  & { 8 }  & { 9 }  & {10} \tabularnewline
\hline 
1  & C/1972 L1  & 4.18 (3.59)  & 4.26  & -0.20\quad{}-0.09\quad{}+0.81  & 61.9  & 91.3\degr  & 115.2\degr  & 40.3\degr  & escaping \tabularnewline
5  & C/1976 D2  & 5.50 (5.08)  & 6.88  & -2.12\quad{}-1.38\quad{}-0.88  & 99.9  & 120.4\degr  & 126.1\degr  & 44.5\degr  & returning \tabularnewline
6  & C/1976 U1  & 5.74 (7.00)  & 5.86  & -0.34\quad{}-0.13\quad{}50.4  & 51.5  & 53.8\degr  & 90.6\degr  & 30.5\degr  & returning \tabularnewline
9  & C/1979 M3  & 4.40 (3.22)  & 4.69  & -0.39\quad{}-0.29\quad{}19.7  & 59.5  & 304.8\degr  & 337.1\degr  & -14.2\degr  & escaping \tabularnewline
10  & C/1980 E1  & 2.16 (2.28)  & 3.17  & -1.35\quad{}-1.01\quad{}-0.76  & 100  & 163.8\degr  & 164.8\degr  & 13.3\degr  & escaping \tabularnewline
22  & C/1997 J2  & 2.80 (2.99)  & 3.05  & -0.27\quad{}-0.25\quad{}-0.23  & 100  & 179.5\degr  & 179.5\degr  & 0.45\degr  & escaping \tabularnewline
\hline
\end{tabular}} 
\end{table*}

\par\end{center}

The remaining three comets with extremely large past semimajor axes
are C/1978~G2, C/2004~X3 and C/2005~K1. All three have mixed past
swarms of VCs but while C/2005~K1 and C/2004 X3 have all VCs in elliptical
orbits, the majority of VCs representing C/1978~G2 are hyperbolic.
However it must be noted, that C/1978~G2 has poorly determined orbit.
In fact it is the worst determined orbit throughout our sample, what
comes from an extremely small number of observations - we have only
7 positions of that comet. As a result, its past swarm of VCs is very
dispersed. Nevertheless, it is the only one comet in our sample with
$1/a_{{\rm ori}}$ formally negative. Having 72\% of past VCs (including
a nominal one) moving on hyperbolic orbits this comet seems to be
a candidate for an interstellar one.

The past and future dynamical evolution of C/2005~K1 is shown in
Fig.~\ref{fig:fon_2005k1x}. We plot here the heliocentric distance
of all 5001 VCs with respect to the time. The zero point in the time
axis corresponds to the observed perihelion passage of this comet
(2005 Nov. 21). To obtain such a plot for this particular comet, we
allowed all VCs to move as far as 500\,000\,au from the Sun, what
takes more than 80 million years for some of them. Even such an extremely
distant escape limit is not sufficient to obtain purely returning
past swarm of VCs for this comet. While all orbits are elliptical,
their high dispersion forced us to conclude, that the dynamical history
of C/2005~K1 cannot be determined basing on available observations.
In contrast, all its future VCs are ejected from the Solar System
in hyperbolic orbits (nominal $1/a_{{\rm fut}}=(-82.4\pm3.3)\times10^{-6}$au$^{-1}$)
without any doubt. There is an additional interesting detail in its
past evolution depicted in Fig.~\ref{fig:fon_2005k1x}. We marked
with black lines the future motion of the nominal VC (a-curve), the
past motion of the nominal VC (b-curve) and the past motion of one
additional VC (c-curve), which represents an interesting dynamical
scenario. Its orbital period is equal to the period of long term perihelion
distance changes due to the Galactic tides. As a result its previous
perihelion distance can be arbitrarily small. However, because it
takes some 57 million years to reach previous perihelion for this
VC, one should treat this scenario as rather questionable due to potential
stellar perturbations suffered by it during such a long time interval.

The past evolution of VC~swarm of C/2004~X3 is much more tight\emph{
}than that of comets C/1978~G2 and C/2005~K1. However, to obtain
the majority of clones coming back it was necessary to shift the escape
limit to 140\,000\,au (see Table~\ref{tab:past_motion_new}) and
only unrealistically large escape limit of 260\,000\,au gives a
whole VC~swarm returning.

\begin{figure}
\includegraphics[height=1\columnwidth,angle=-90]{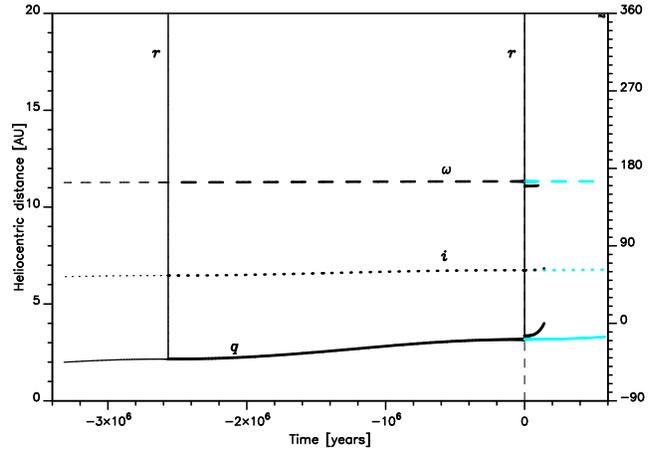}

\caption{\label{fig:gal_1980e1}Past and future orbital evolution of the nominal
VC for C/1980~E1, see text for a detailed description. This comet
passed previous perihelion 2.6 Myr ago at 2~au and it will be ejected
in the future in hyperbolic orbit. The effect of strong planetary
perturbation is easily visible .}

\end{figure}

\subsection{Small previous perihelion passage distances}

\label{sub:Small-previous-perihelion-passage}

In the light of Jupiter-Saturn barrier concept one should expect,
that most of the Oort spike comets should have the previous perihelion
distance well out of rich of planetary perturbations. This is not
the case. In the sample of 22 small perihelion comets ($q_{{\rm osc}}<3.0$\,au)
investigated in Paper I we obtained previous perihelion distances
$q_{{\rm prev}}<15$\,au for 15~comets (almost 70 per cent). Now,
in the sample of 64~LPCs with $q_{{\rm osc}}>3$\,au we obtained
significantly smaller fraction of such comets, but still almost 50
per cent of the sample have the previous perihelion distance smaller
than 15\,au. Moreover, among them, 6~comets (C/1972~L1, C/1976~D2,
C/1976~U1, C/1979~M3, C/1980~E1 and C/1997~J2) have the previous
perihelion distance smaller than the osculating one (see Table~\ref{tab:q_increase})!
When comparing with Paper I, the percentage of comets observed at
the greater perihelion distance than the previous one is roughly the
same, but the $q$-changes for small perihelion comets are smaller,
below 0.3\,au. It should be stressed here, that observing LP~comets
during the increasing phase of their perihelion distance evolution
is the direct evidence that their dynamical history followed one of
the two possibilities: either they were strongly perturbed by planets
during their previous perihelion passage (what switched the phase
of the perihelion distance evolution) or they have moved unperturbed
through the Jupiter-Saturn barrier in the past. Of these six comets,
C/1976~D2 suffered practically no planetary perturbations in the
observed perihelion passage (see Fig.~\ref{fig:gal_1976d2}). These
comets can also be found in the very bottom part of Fig.~\ref{fig:gal_deltaq}
with some theoretical interpretation of this distribution given in
Section \ref{sub:Evolution-of-orbital-elements-in-Galactic-frame}.

An example of such a situation is depicted in Fig.~\ref{fig:gal_1980e1}
where past and future dynamical evolution of C/1980~E1 is illustrated
by the evolution of orbital elements of its nominal VC. Like for any
other VC we followed numerically its motion under the influence of
Galactic perturbation. Since we present several similar plots, we
describe it here in more detail. For the past motion we started from
the \emph{original} orbit while for the future motion from\emph{ future}
orbit, both were obtained with the NG~effects included in this specific
case, see Section~\ref{sub:Non-gravitational-forces-detection} for
additional information. The horizontal axis shows the moment of osculation,
for which orbital elements are calculated and plotted, the zero point
corresponds to the observed perihelion passage. The left vertical
axis is expressed in au and describes both heliocentric distance of
a VC ($r$, thin vertical lines) and its perihelion distance ($q$,
continuous line). The right vertical axis describes angular elements
(calculated in the Galactic frame) and is expressed in degrees. We
plot here the synchronous evolution of the argument of perihelion
($\omega$, dashed line) and inclination ($i$, dotted line). The
thick lines depict the real dynamical VC~evolution while their continuation
with the thin lines depicts its potential motion in the absence of
planetary perturbations in the previous/next perihelion. All grey
(cyan) lines (right from the zero point) describe additionally an
artificial variant of the future motion, in the absence of all planetary
perturbations during the observed perihelion passage too. The discontinuities
of the thick lines at the zero point of the time axis are the result
of a close encounter of C/1980~E1 with Jupiter ( $\Delta=0.228$~au,
9.46 December 1980).

The original perihelion distance of C/1980~E1 $q_{{\rm ori}}=$3.17\,au
while the previous one (almost 2.6\,million years ago) $q_{{\rm prev}}=2.16$\,au.
It should be noted, that at previous perihelion passage C/1980~E1
was perturbed by planets, possibly rather strongly due to its small
previous perihelion distance. Since it is impossible to calculate
this perturbation (due to large uncertainties in planetary positions
2.6\,million years ago) one should treat the orbital element evolution
left from the previous perihelion passage likely to be completely
fictitious so we pointed them in thin lines.

The largest perihelion distance increase one can found for C/1976~D2,
where the observed value was 6.88~au but the previous one 5.50~au
(see Table~\ref{tab:q_increase}). This comet is also a clear evidence,
that moving through the Jupiter-Saturn barrier can be observed. Planetary
perturbations in this case were very weak, slightly decreasing its
inverse semimajor axis from $1/a_{{\rm ori}}=(56.9\pm7.3)\times10^{-6}$\,au\,$^{-1}$
to $1/a_{{\rm fut}}=(54.1\pm7.3)\times10^{-6}$\,au\,$^{-1}$. This
almost unperturbed motion through perihelion is depicted in Fig.\ref{fig:gal_1976d2}.
In fact, this comet is a \emph{'double evidence'}. First, we observed
it at larger perihelion distance than the previous one, with all consequences
described above. Second, the observed perihelion passage demonstrated
an unperturbed motion through the Solar System and its next perihelion
distance is even larger, while still observable! This is discussed
in a more detail in the next section.

\begin{figure}
\includegraphics[bb=70bp 26bp 586bp 761bp,height=1\columnwidth,angle=-90]{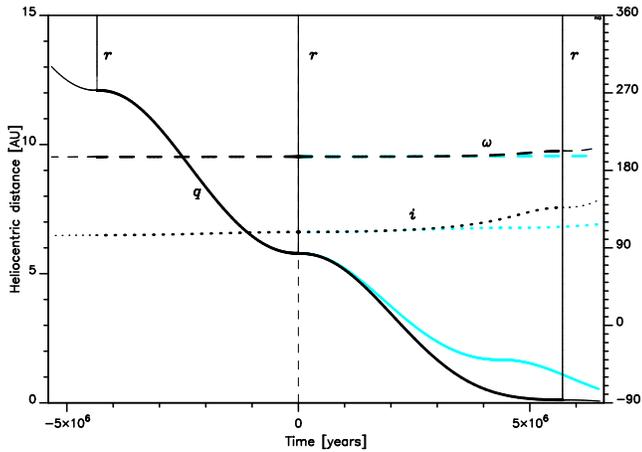}

\caption{\label{fig:gal_1999f1}Past and future orbital evolution of the nominal
VC for C/1999~F1, see text for a detailed description. Another example
of Oort spike comet moving through the Jupiter-Saturn barrier and
remaining observable LP~comet. 50 per cent of VCs representing this
comet will have the next perihelion distance smaller than 0.012~au
and 90 per cent smaller then 0.27~au.}

\end{figure}

\subsection{Small next perihelion passage distances}

\label{sub:Small-next-perihelion-passage}

While the Jupiter-Saturn barrier mechanism predicts that great majority
of all LP comets that approach the Sun closer than 10--15\,au should
be definitely removed from this population, we observe that over 40
per cent of our sample ( 26~comets ) will keep moving on the typical
LP comet orbits with small ($q<10$\,au) next perihelion passage
distances. Moreover, six of them remain to be members of the Oort
spike ($1/a_{{\rm fut}}<10^{-4}$\,au). These comets (C/1976~D2,
C/1999~F1, C/2000~A1, C/2002~L9, C/2004~T3 and C/2005~G1) constitute
an important, direct evidence, that about 10 per cent of the large
perihelion distance Oort spike comets can move directly through the
Jupiter-Saturn barrier and remain observable. It is worth to mention
that for a hypothetical observer of such comets next perihelion passage
they can successfully pretend to be a result of the \citet{kaib-quinn:2009}
scenario discussed in Section~ \ref{sub:New-interpretations}. Three
of them have their semimajor axes significantly shortened, what makes
the perihelion distance evolution under the Galactic tides much slower.

An impressive example of very small next perihelion distance is the
case of C/1999~F1, see Fig.~\ref{fig:gal_1999f1}. The previous
perihelion distance of this comet was $12.1\pm0.48$\,au, the observed
one 5.79\,au but the next with high certainty will be smaller than
0.3\,au! In contrary to the most probable scenario attributed to
the Jupiter-Saturn barrier crosser, the semimajor axis of this comet
was slightly increased due to the planetary perturbations (the same
happened to C/1976~D2) and as a result the next perihelion passage
will be closer to the Sun than in the absence of planets. Typical
planetary perturbation here are very small. In Paper~I we obtained
only 9~returning comets for the future motion (from the sample of
22~comets) but all of them have perihelia inside the observable zone.

\subsection{Comets with next semimajor axis below 2\,000\,au}

\label{sub:Small-next-semimajor-axis}

According to our analysis $\sim12$ per cent (8 objects) of the observed
large perihelion Oort spike comets return in orbits similar to that
of comet C/1996~B2 Hyakutake ($(1/a)_{{\rm fut}}=554\times10^{-6}$\,au$^{-1}$),
however that comet had the original orbit more tightly bound than
future orbit. Six of these comets creates a noticeable local maximum
in the $1/a_{{\rm fut}}$ distribution displayed in Fig.~\ref{fig:distribution_a_all}.
This maximum consists of two dynamically new comets, three dynamically
old comets and one with an uncertain past dynamical status (see the
definitions in Section~\ref{sub:How-we-distinguish}). Three of these
objects have $q_{{\rm osc}}<3.5$\,au. The shortest future orbit is for
C/2002~A3, which future semimajor axis equals $\sim162$~au. This
comet will return in about 1600 years with the perihelion distance
of 5.15\,au.

\begin{center}

\begin{table}
\caption{\label{tab:sample_q_dist} Observed perihelion distribution in the
sample of large perihelion Oort spike comets (for explanation of dynamically
new, dynamically old and dynamically uncertain comets see Section~\ref{sec:New-and-old}).}

\centering{}\begin{tabular}{ccccccc}
\hline 
$q$ {[}au{]}  & 3--4  & 4--5  & 5--6  & 6--7  & 7--10  & all \tabularnewline
\hline 
all  & 23  & 12  & 14  & 8  & 7  & 64 \tabularnewline
\hline 
dynamically new comets  & 10  & 6  & 5  & 5  & 5  & 31 \tabularnewline
dynamically old comets  & 13  & 4  & 7  & 2  & 0  & 26 \tabularnewline
dynamically uncertain comets  & 0  & 2  & 2  & 1  & 2  & 7 \tabularnewline
\hline
\end{tabular}
\end{table}

\par\end{center}

\begin{center}

\begin{table}
\caption{\label{tab:sample_i_dist} Observed inclination distribution in the
sample of large perihelion Oort spike comets (for explanation of dynamically
new, dynamically old and dynamically uncertain comets see Section~\ref{sec:New-and-old}).}

\centering{}\begin{tabular}{cccccc}
\hline 
$i$ {[}deg{]}  & \multicolumn{2}{c}{$i<90$\degr} & \multicolumn{2}{c}{$i\geq90$\degr} & all \tabularnewline
$q$ {[}au{]}  & 3--4.5  & 4.5--10  & 3--4.5  & 4.5--10  & \tabularnewline
\hline 
all  & 14  & 22  & 14  & 14  & 64 \tabularnewline
\hline 
dynamically new comets  & 9  & 13  & 4  & 5  & 31 \tabularnewline
dynamically old comets  & 5  & 5  & 10  & 6  & 26 \tabularnewline
dynamically uncertain comets  & --  & 4  & --  & 3  & 7 \tabularnewline
\hline
\end{tabular}
\end{table}

\par\end{center}

\section{New and old LP comets}

\label{sec:New-and-old}

The terms of \emph{new} and \emph{old }long-period comets are widely
used in literature for many decades. Sometimes the authors add adjectives:
\emph{dynamically} or \emph{physically}, to inform the reader, what
criteria they use to distinguish between \emph{new} and \emph{old
}LPCs, but in most cases, the intention is that \emph{dynamically
new} should appear as \emph{physically new} and vice verse. Historically,
the first criterion used in this field was simply the semimajor axis,
$a$, value. The widely accepted statement was, that all comets with
$a>$10\,000\,au were dynamically new. This was used for example
by Oort \citeyearpar{oort:1950} when introducing his concept of the
distant comet reservoir, now called the Oort Cloud. Just a year later,
\citet{oort-schmidt:1951} published a paper which seems to be the
source of the widely quoted and repeated opinion, that new LPCs are
more active and brighter. In fact, nowadays it is very difficult to
prove the truth of this statement, see for example \citet{dyb-hist:2001},
who collected large number of LPCs absolute magnitudes and found no
correlation with their dynamical history. \citet{dyb-hist:2001} also
showed, that using $a>$10\,000\,au as the criterion of being the
\emph{dynamically new} LPC\emph{ }seems to be completely unsatisfactory.

Recently, \citet{fink:2009} presented an extended taxonomic survey
of comet composition, based on their spectroscopic observations. As
it concerns LP~comets, he also did not found any correlation with
the semimajor axis (see for example Fig.~8 in the quoted paper).

\begin{figure}[t]
\includegraphics[height=1\columnwidth,angle=-90]{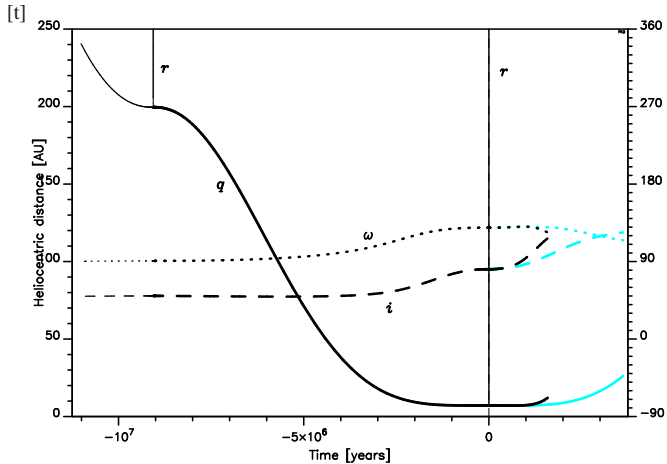}

\caption{\label{fig:gal_1999j2}Past and future orbital evolution of the nominal
VC for C/1999~J2. An example of dynamically new comet; from the swarm
of its VCs it comes, that 90 per cent of them had their previous perihelion
distance greater than 160\,au.}

\end{figure}

\begin{figure}[t]
\begin{centering}
\includegraphics[width=8.6cm]{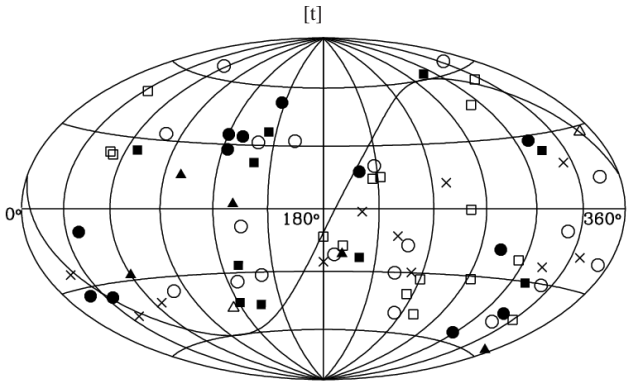} 
\par\end{centering}

\caption{This Aitoff projection sky map shows the distribution of aphelion
directions for all (discovered before 2009) large perihelion distance
Oort spike comets in galactic coordinates. Squares, circles and triangles
show comets investigated in this paper where symbols indicate dynamically
old comets, dynamically new and comets of uncertain dynamical status,
respectively; full and open markers represent comets returning and
escaping in the future. Black crosses show 11 comets discovered before
1970. }

\centering{}\label{fig:aitoff_galactic} 
\end{figure}

\subsection{\label{sub:How-we-distinguish}How we distinguish them?}

Let's start with definitions. We use the term \emph{dynamically old}
LP~comet for objects with \emph{previous} perihelion passage distance
smaller than some threshold. This threshold value should describe
the sphere of significant planetary perturbations. In Paper~I we
used 15\,au as this limit but now we decided to use three different
values, namely 10, 15 and 20\,au in parallel to observe how the \emph{new/old}
classification depends on it for investigated comets. Because we replaced
each individual comet with a swarm of 5001~VCs we applied the above-mentioned
criterion individually to each VC and then classified a particular
comet depending on the percentage of escaping VCs -- if more than
50 per cent of VCs were escaping in the past we call the parent comet
a \emph{dynamically new} one with respect to the particular threshold
value. It is worth to mention, that except of a few cases this percentage
is significantly closer to zero or 100 per cent, see col. {[}11{]}
in Tables~\ref{tab:past_motion_old} -- \ref{tab:past_motion_uncertain}.
With this definition, \emph{a dynamically new} LP~comet should have
moved (before the observed perihelion passage) in the orbit which
is free from planetary perturbations and therefore can be used to
study the source region of LP~comets by tracing its motion back in
time under the Galactic perturbations. From the point of view of the
above-mentioned three different threshold values and basing on their
motion in the past we finally divided the whole sample of 64~comets
into three groups. In the first one, see Table~\ref{tab:past_motion_old},
we placed 26 comets \emph{dynamically old} with respect to all three
values. The second group, see Table~\ref{tab:past_motion_new}, consists
of 31 \emph{dynamically new} comets with respect to all three values
(26~comets) or only to the two lower values (10\,au and 15\,au;
5 comets.) The third one, see Table~\ref{tab:past_motion_uncertain},
groups 7 comets of the uncertain dynamical age: they are \emph{new}
if one take 10\,au as the threshold value, but \emph{old }for the
greater threshold values.

\begin{figure*}
\begin{centering}
\includegraphics[width=5.73cm]{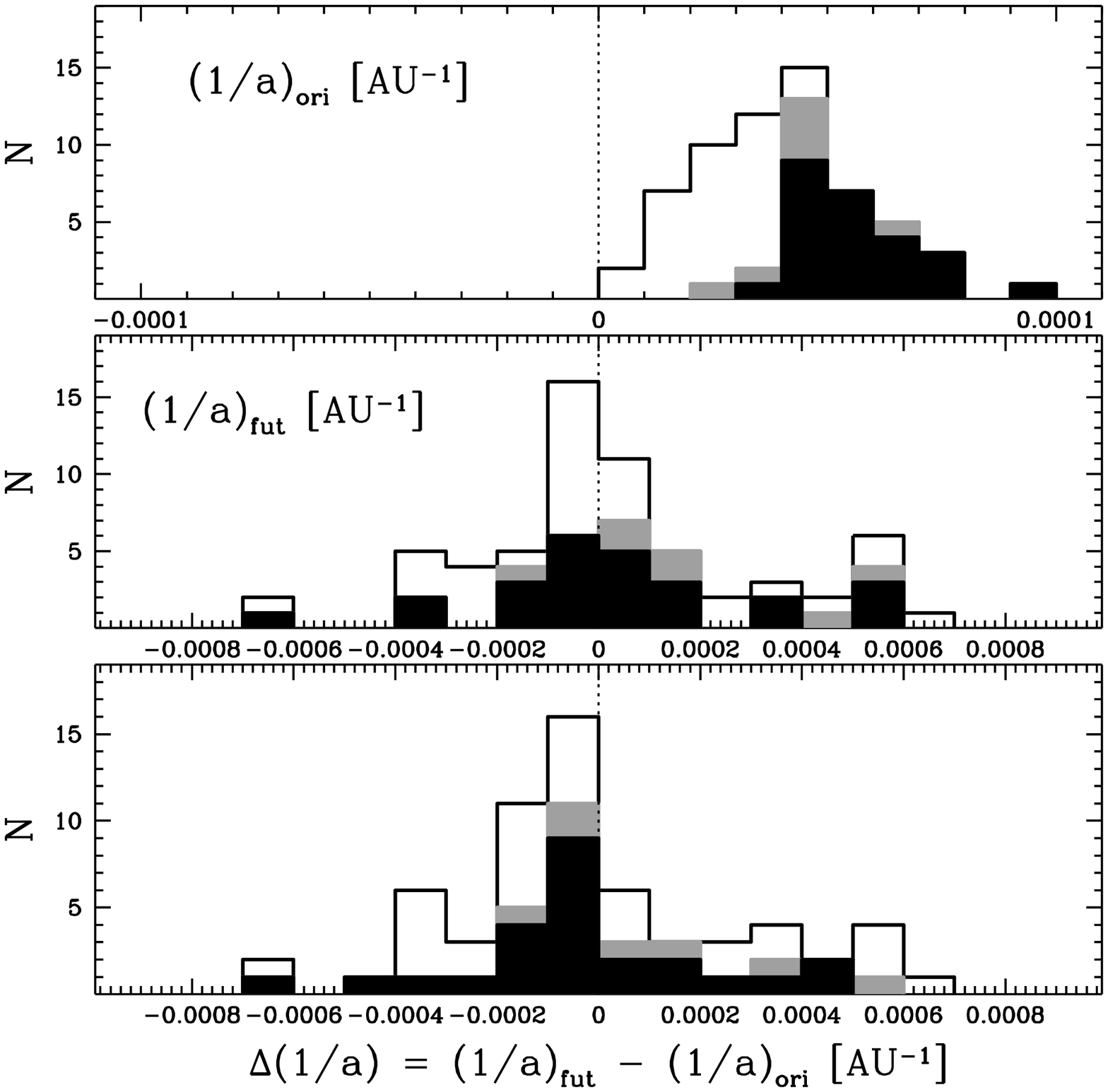} \includegraphics[width=5.73cm]{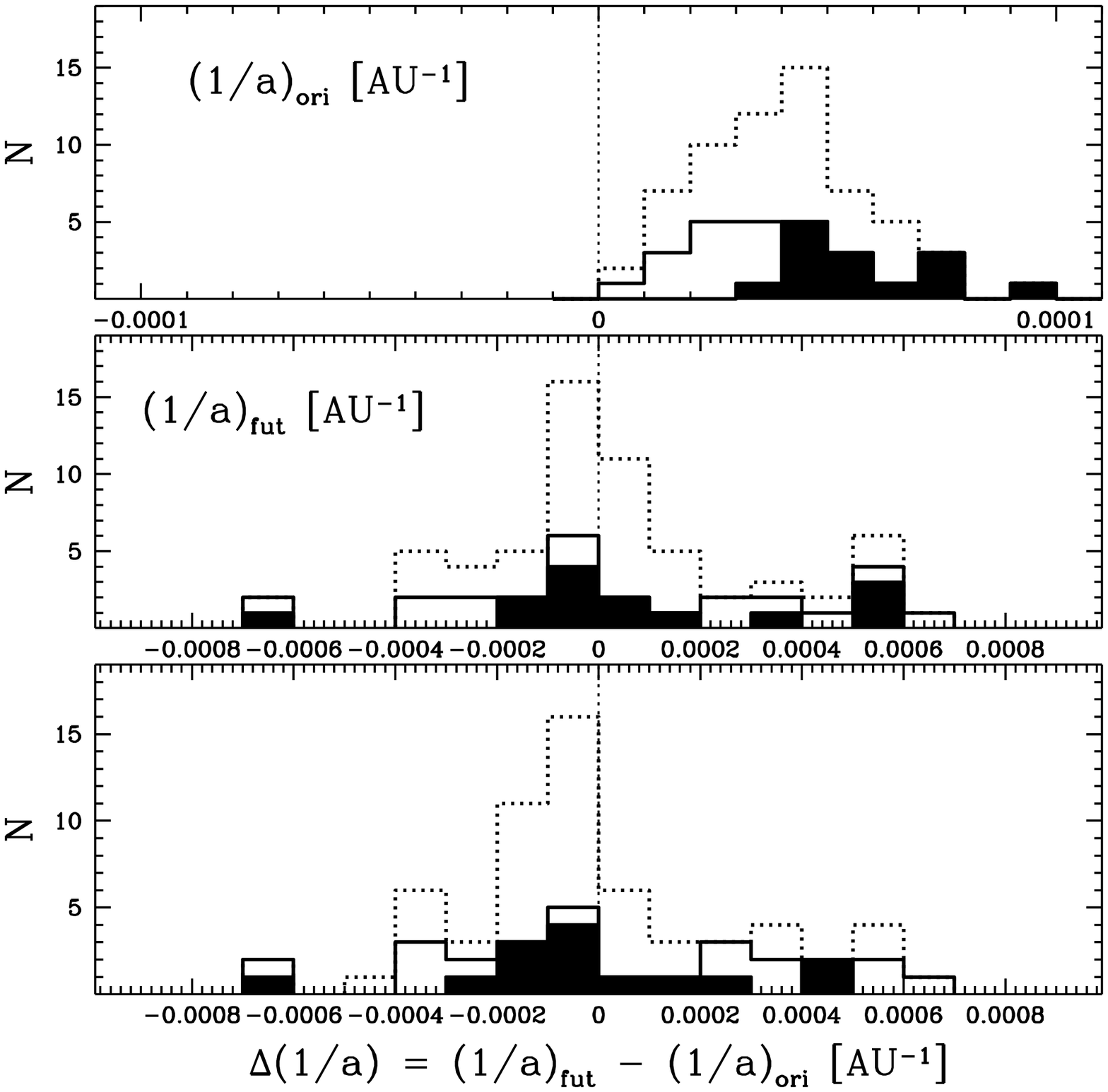}
\includegraphics[width=5.73cm]{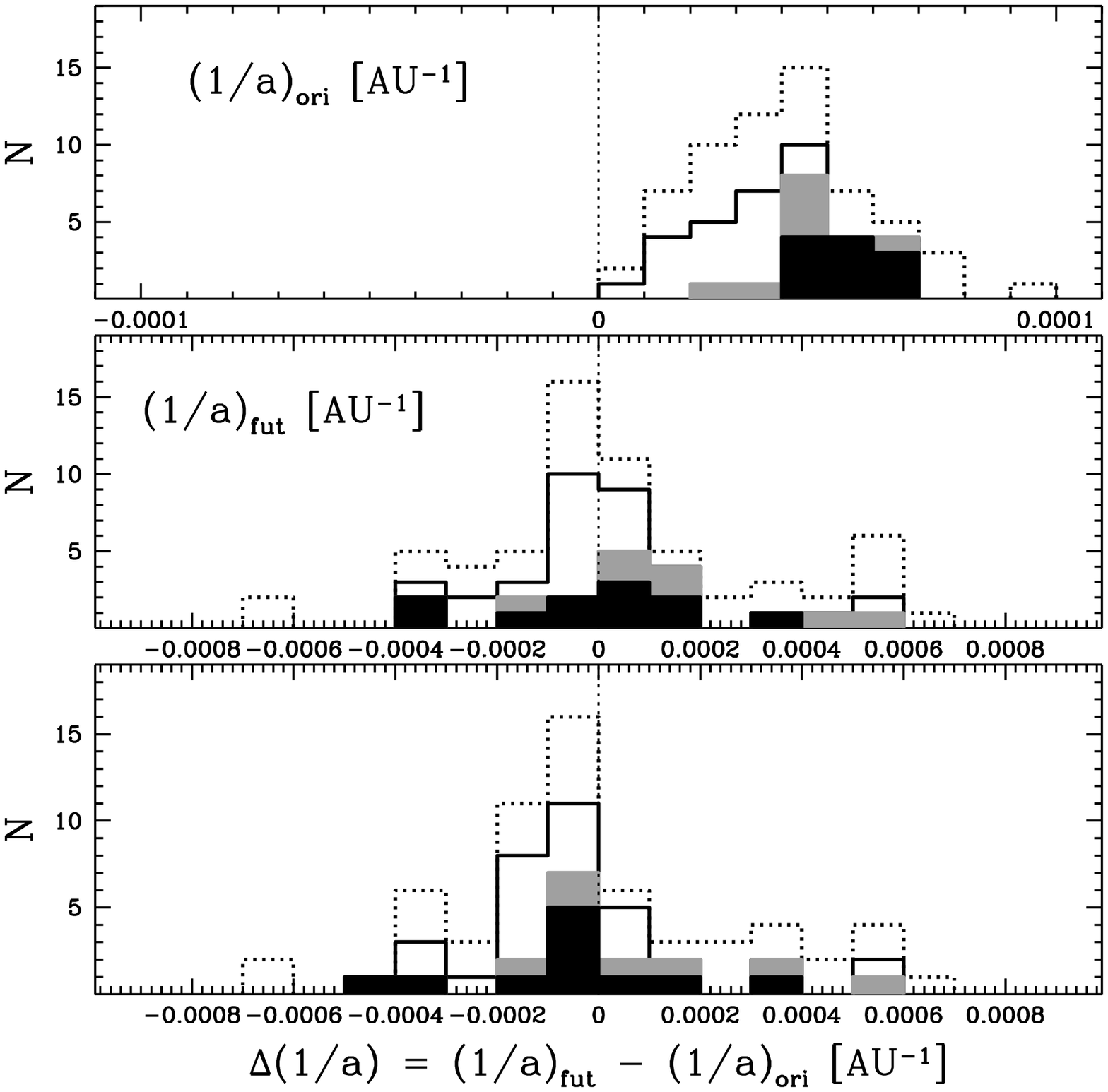} 
\par\end{centering}

\begin{centering}
(a) $q_{{\rm osc}}\geq3.0$\,au \hspace{3.7cm} (b) 3.0\,$\leq q_{{\rm osc}}<4.5$\,au
\hspace{3.7cm} (c) $q_{{\rm osc}}\geq4.5$\,au 
\par\end{centering}

\caption{Distributions of original and future cometary energies measured by
$1/a_{{\rm ori}}$ (top panels) and $1/a_{{\rm fut}}$
(middle panels) for the observed sample of large perihelion Oort~spike
comets. The bottom panels represent the distribution of planetary
perturbations acting on the comets during their passage through the
planetary system ($\Delta(1/a)=1/a_{{\rm fut}}-1/a_{{\rm ori}}$).
The filled parts of histograms represent the dynamically old comets,
and grey - the seven comets with uncertain dynamical status, i.e.
comets not dynamically new for the limit of 15\,au for previous perihelion
distance but being dynamically new for the limit of 10\,au. Dotted
histograms in the middle and right-hand side panels ((b) and (c))
represent the total histograms of 64~comets from the left-hand side
panel (a).}

\centering{}\label{fig:distribution_a_all} 
\end{figure*}

\begin{table*}
\caption{\label{tab:future_motion_returning}The future distributions of the
returning and mixed swarms of VCs in terms of returning {[}R{]}, escaping {[}E{]},
including hyperbolic {[}H{]} VC numbers. Aphelion and perihelion distances
are described either by a mean value for the normal distributions,
or three deciles at 10, 50 (i.e. median), and 90 per cent. In the
case of mixed swarm the mean values or deciles of $Q$ and $q$ are
given for the returning part of the VCs swarm, where the escape limit
of 120\,000\,au was generally adopted with one exception of comet
C/2002~J5, marked with an asterisk, where the escape limit of 140\,000\,au
was applied. The upper-a index in columns 4-5 means that this part
of mixed swarm includes the nominal orbit. Last column presents the
value of $1/a_{{\rm fut}}$.}

\begin{tabular}{rlccccccr@{$\pm$}r@{}}
\hline 
\#  & Comet  & \multicolumn{3}{c}{Number of VCs} & eccentricity  & $Q_{{\rm next}}$  & $q_{{\rm next}}$  & \multicolumn{2}{c}{$1/a_{{\rm fut}}$}\tabularnewline
 &  & {[}R{]}  & {[}E{]}  & {[}H{]}  &  & $10^{3}$au  & au  & \multicolumn{2}{c}{$10^{-6}$au$^{-1}$}\tabularnewline
{[}1{]}  & {[}2{]}  & {[}3{]}  & {[}4{]}  & {[}5{]}  & {[}6{]}  & {[}7{]}  & {[}8{]}  & \multicolumn{2}{c}{{[}9{]}}\tabularnewline
\hline 
{ 3 }  & {C/1974 F1$^{{\rm NG}}$}  & 5001  & 0  & 0  & 0.998429 $\pm$ 0.000020  & 3.829 $\pm$ 0.050  & 3.009926 $\pm$ 0.000019  & 522.1  & 6.8 \tabularnewline
{ 4 }  & C/1974 V1  & 5001  & 0  & 0  & 0.996545 $\pm$ 0.000073  & 3.476 $\pm$ 0.073  & 6.012081 $\pm$ 0.000084  & 574.4  & 12.1 \tabularnewline
{ 5 }  & C/1976 D2  & 5001  & 0  & 0  & 0.999443 - 0.999489 - 0.999496  & 31.5 - 36.9 - 44.9  & 8.2 - 9.3 - 12.4  & 54.1  & 7.3 \tabularnewline
{ 6 }  & C/1976 U1  & 5001  & 0  & 0  & 0.99884 $\pm$ 0.00013  & 8.8 - 10.1 - 11.8  & 5.8669 $\pm$ 0.0038  & 197.1  & 21.9 \tabularnewline
{ 8 }  & C/1978 G2  & 7  & 4994  & 4984$^{a}$  & 0.9907 - 0.9969 - 0.9998{[}R{]}  & 49.6 - 87.6 - 105.4{[}R{]}  & 6.07 - 134.4 - 483.0{[}R{]}  & -99.2  & 37.8 \tabularnewline
{13 }  & C/1987 F1  & 5001  & 0  & 0  & 0.999374 $\pm$ 0.000018  & 11.48 $\pm$ 0.33  & 3.5912 $\pm$ 0.0035  & 174.2  & 5.0 \tabularnewline
{17 }  & C/1992 J1  & 5001  & 0  & 0  & 0.9983576 $\pm$ 0.0000028  & 3.655 $\pm$ 0.006  & 3.004545 $\pm$ 0.000005  & 546.7  & 0.9 \tabularnewline
{19 }  & C/1993 K1  & 5001  & 0  & 0  & 0.997118 $\pm$ 0.000037  & 3.358 $\pm$ 0.043  & 4.845977 $\pm$ 0.000021  & 594.8  & 7.7 \tabularnewline
{21 }  & {C/1997 BA$_{6}^{{\rm NG}}$}  & 5001  & 0  & 0  & 0.9981686 $\pm$ 0.0000059  & 4.966 $\pm$ 0.021  & 3.432091 $\pm$ 0.000006  & 402.5  & 1.7 \tabularnewline
{22 }  & {C/1997 J2$^{{\rm NG}}$}  & 151  & 4850$^{a}$  & 0  & 0.9999168 - 0.9999173 - 0.9999176{[}R{]}  & 117 - 120 - 121{[}R{]}  & 4.81 - 4.94 - 5.04{[}R{]}  & 14.7  & 0.9 \tabularnewline
{23 }  & C/1999 F1  & 5001  & 0  & 0  & 0.9999915 - 0.9999965 - 0.9999988  & 64.2 $\pm$ 1.4  & 0.040 - 0.114 - 0.266  & 31.2  & 0.7 \tabularnewline
{24 }  & C/1999 F2  & 5001  & 0  & 0  & 0.998348 $\pm$ 0.000016  & 57.170 $\pm$ 0.054  & 4.724284 $\pm$ 0.000023  & 349.5  & 3.3 \tabularnewline
{27 }  & C/1999 K5  & 5001  & 0  & 0  & 0.9987481 $\pm$ 0.0000031  & 5.209 $\pm$ 0.013  & 3.2628366 $\pm$ 0.0000047  & 383.7  & 0.9 \tabularnewline
{28 }  & C/1999 N4  & 5001  & 0  & 0  & 0.99743 - 0.99864 - 0.99928  & 77.2 - 84.2 - 92.3  & 27.7 - 57.3 118.7  & 23.8  & 1.7 \tabularnewline
{30 }  & C/1999 U1  & 5001  & 0  & 0  & 0.997780 $\pm$ 0.000012  & 3.716 $\pm$ 0.021  & 4.129871 $\pm$ 0.000023  & 537.7  & 3.0 \tabularnewline
{32 }  & {C/1999 Y1$^{{\rm NG}}$}  & 5001  & 0  & 0  & 0.9989332 $\pm$ 0.0000045  & 5.783 $\pm$ 0.024  & 3.08617 $\pm$ 0.00003  & 345.7  & 1.5 \tabularnewline
{33 }  & C/2000 A1  & 5001  & 0  & 0  & 0.999375 $\pm$ 0.000026  & 26.87 $\pm$ 0.72  & 8.38 $\pm$ 0.12  & 74.5  & 2.0 \tabularnewline
{34 }  & {C/2000 CT$_{54}^{{\rm NG}}$}  & 5001  & 0  & 0  & 0.9981542 $\pm$ 0.0000080  & 3.414 $\pm$ 0.015  & 3.1532274 $\pm$ 0.0000066  & 585.3  & 2.5 \tabularnewline
{35 }  & C/2000 K1  & 5001  & 0  & 0  & 0.999148 $\pm$ 0.000015  & 14.54 $\pm$ 0.25  & 6.1938 $\pm$ 0.0051  & 137.6  & 2.3 \tabularnewline
{36 }  & C/2000 O1  & 5001  & 0  & 0  & 0.999260 $\pm$ 0.000029  & 15.81 $\pm$ 0.60  & 5.829 - 5.842 - 5.853  & 126.6  & 4.7 \tabularnewline
{38 }  & C/2000 Y1  & 1  & 5000$^{a}$  & 1721  & 0.9841{[}R{]}  & 118.3{[}R{]}  & 949.8{[}R{]}  & 1.6  & 4.2 \tabularnewline
{41 }  & C/2001 K3  & 5001  & 0  & 0  & 0.997962 $\pm$ 0.000021  & 3.002 $\pm$ 0.031  & 3.060955 $\pm$ 0.000010  & 665.7  & 6.8 \tabularnewline
{43 }  & C/2002 A3  & 5001  & 0  & 0  & 0.968222 $\pm$ 0.000010  & 0.3188 $\pm$ 0.0001  & 5.14743 $\pm$ 0.00001  & 6173.6  & 1.9 \tabularnewline
{45 }  & C/2002 J5$^{*}$  & 2352  & 2649$^{a}$  & 0  & 0.9668 - 0.9705 - 0.9759{[}R{]}  & 137.8 - 143.7 - 147.3{[}R{]}  & 1681 - 2151 - 2485{[}R{]}  & 13.3  & 0.7 \tabularnewline
{46 }  & C/2002 L9  & 5001  & 0  & 0  & 0.9994906$\pm$ 0.0000072  & 25.7 $\pm$ 0.3  & 6.550 $\pm$ 0.018  & 77.8  & 0.9 \tabularnewline
{47 }  & {C/2002 R3$^{{\rm NG}}$}  & 165  & 4836$^{a}$  & 1180  & 0.999883 - 0.999947 - 1.000051{[}R{]}  & 93.5 - 115.7 - 125.3{[}R{]}  & 214 - 838 - 1381{[}R{]}  & 3.9  & 6.3 \tabularnewline
{50 }  & C/2003 WT$_{42}$  & 5001  & 0  & 0  & 0.9989720 $\pm$ 0.0000017  & 10.014 $\pm$ 0.016  & 5.14993 $\pm$ 0.00018  & 199.6  & 0.3 \tabularnewline
{52 }  & C/2004 T3  & 5001  & 0  & 0  & 0.999418 $\pm$ 0.000048  & 25.0 - 26.8 - 28.9  & 7.52 - 7.82 - 8.04  & 74.6  & 4.2 \tabularnewline
{54 }  & {C/2005 B1$^{{\rm NG}}$}  & 5001  & 0  & 0  & 0.9992321 $\pm$ 0.0000021  & 8.362 $\pm$ 0.023  & 3.211764 $\pm$ 0.000068  & 239.1  & 0.7 \tabularnewline
{56 }  & C/2005 G1  & 5001  & 0  & 0  & 0.99973125 - 0.99973237 - 0.99973261  & 41.79 $\pm$ 0.89  & 5.46 - 5.59 - 5.74  & 47.9  & 1.0 \tabularnewline
{61 }  & {C/2006 S2$^{{\rm NG}}$}  & 310  & 4691  & 3633$^{a}$  & 0.99434 - 0.99874 - 0.99979{[}R{]}  & 56.3 - 86.5 - 114.0{[}R{]}  & 5.60 - 54.5 - 322.3{[}R{]}  & -10.8  & 18.1 \tabularnewline
{62 }  & C/2006 YC  & 5001  & 0  & 0  & 0.997841 $\pm$ 0.000060  & 45.74 $\pm$ 0.13  & 4.940736 $\pm$ 0.000078  & 437.2  & 12.1 \tabularnewline
{64 }  & C/2007 Y1  & 5001  & 0  & 0  & 0.999050 $\pm$ 0.000042  & 7.04 $\pm$ 0.31  & 3.34178 - 3.34202 - 3.34221  & 284.3  & 12.5 \tabularnewline
\hline
\end{tabular}
\end{table*}

\begin{table*}
\caption{\label{tab:future_motion_escaping}The future distributions of comets
with fully hyperbolic {[}H{]} swarms of VCs. Comets with NG effects
are indicated by upper-NG index located behind the comet designation
(column 2). In column~6 the eccentricity distribution at 120\,000\,au
are described either by a mean value for the normal distributions,
or three deciles at 10, 50 and 90 per cent. Last column presents the
value of $1/a_{{\rm fut}}$. }

\begin{tabular}{rlccccr@{$\pm$}r@{}}
\hline 
\#  & Comet  & \multicolumn{3}{c}{Number of VCs} & eccentricity  & \multicolumn{2}{c}{$1/a_{{\rm fut}}$}\tabularnewline
 &  & {[}R{]}  & {[}E{]}  & {[}H{]}  & at 120\,000 au  & \multicolumn{2}{c}{$10^{-6}$au$^{-1}$}\tabularnewline
{[}1{]}  & {[}2{]}  & {[}3{]}  & {[}4{]}  & {[}5{]}  & {[}6{]}  & \multicolumn{2}{c}{{[}7{]}}\tabularnewline
\hline 
{ 1 }  & C/1972 L1  & 0  & 5001  & 5001  & 1.003972 $\pm$ 0.000057  & -620.9  & 6.2 \tabularnewline
{ 2 }  & C/1973 W1  & 0  & 5001  & 5001  & 1.000270 - 1.000342 - 1.000429  & -107.0  & 12.3 \tabularnewline
{ 7 }  & C/1978 A1  & 0  & 5001  & 5001  & 1.0001178 $\pm$ 0.0000061  & -97.4  & 11.9 \tabularnewline
{ 9 }  & C/1979 M3  & 0  & 5001  & 5001  & 1.000999 $\pm$ 0.000130  & -144.9  & 14.5 \tabularnewline
{10 }  & {C/1980 E1$^{{\rm NG}}$}  & 0  & 5001  & 5001  & 1.067769 $\pm$ 0.000014  & -16011  & 3\tabularnewline
{11 }  & {C/1983 O1$^{{\rm NG}}$}  & 0  & 5001  & 5001  & 1.0000366 $\pm$ 0.0000031  & -186.9  & 2.1 \tabularnewline
{12 }  & {C/1984 W2$^{{\rm NG}}$}  & 0  & 5001  & 5001  & 1.000165 - 1.000280 - 1.000422  & -31.6  & 8.6 \tabularnewline
{14 }  & C/1987 H1  & 0  & 5001  & 4896  & 1.00000028 -1.00000100 - 1.00000238  & -5.7  & 2.8 \tabularnewline
{15 }  & C/1987 W3  & 0  & 5001  & 5001  & 1.001047 $\pm$ 0.000027  & -361.5  & 7.3 \tabularnewline
{16 }  & C/1988 B1  & 0  & 5001  & 5001  & 1.000682 $\pm$ 0.000070  & -109.3  & 7.0 \tabularnewline
{18 }  & C/1993 F1  & 0  & 5001  & 5001  & 1.0004509 $\pm$ 0.0000011  & -355.9  & 6.3 \tabularnewline
{20 }  & C/1997 A1  & 0  & 5001  & 5001  & 1.001787 $\pm$ 0.000021  & -227.4  & 1.7 \tabularnewline
{25 }  & {C/1999 H3$^{{\rm NG}}$}  & 0  & 5001  & 5001  & 1.000071 $\pm$ 0.000012  & -10.0  & 1.1 \tabularnewline
{26 }  & C/1999 J2  & 0  & 5001  & 5001  & 1.001221 $\pm$ 0.000014  & -87.0  & 0.6 \tabularnewline
{29 }  & C/1999 S2  & 0  & 5001  & 5001  & 1.0007212$\pm$ 0.0000010  & -321.4  & 3.8 \tabularnewline
{31 }  & C/1999 U4  & 0  & 5001  & 5001  & 1.00013459 $\pm$ 0.00000043  & -293.2  & 0.5 \tabularnewline
{37 }  & {C/2000 SV$_{74}^{{\rm NG}}$}  & 0  & 5001  & 5001  & 1.0001075$\pm$ 0.0000036  & -54.9  & 0.6 \tabularnewline
{39 }  & C/2001 C1  & 0  & 5001  & 5001  & 1.001797 $\pm$ 0.000024  & -214.5  & 2.1 \tabularnewline
{40 }  & C/2001 G1  & 0  & 5001  & 5001  & 1.000848 $\pm$ 0.000029  & -104.7  & 2.8 \tabularnewline
{42 }  & C/2001 K5  & 0  & 5001  & 5001  & 1.0011154 $\pm$ 0.0000079  & -94.9  & 0.4 \tabularnewline
{44 }  & C/2002 J4  & 0  & 5001  & 5001  & 1.00003251-1.00003252-1.00003256  & -269.9  & 1.4 \tabularnewline
{48 }  & C/2003 G1  & 0  & 5001  & 5001  & 1.0021234$\pm$ 0.0000048  & -372.7  & 0.6 \tabularnewline
{49 }  & C/2003 S3  & 0  & 5001  & 5001  & 1.0000039-1.0000087-1.0000122  & -6.1  & 2.9 \tabularnewline
{51 }  & C/2004 P1  & 0  & 5001  & 5001  & 1.000284 $\pm$ 0.000011  & -88.9  & 3.0 \tabularnewline
{53 }  & C/2004 X3  & 0  & 5001  & 5001  & 1.006305 $\pm$ 0.000033  & -639.5  & 2.1 \tabularnewline
{55 }  & {C/2005 EL$_{173}^{{\rm NG}}$}  & 0  & 5001  & 5001  & 1.0000472 $\pm$ 0.0000039  & -19.1  & 0.9 \tabularnewline
{57 }  & {C/2005 K1$^{{\rm NG}}$}  & 0  & 5001  & 5001  & 1.000970 $\pm$ 0.000058  & -82.4  & 3.3 \tabularnewline
{58 }  & C/2005 Q1  & 0  & 5001  & 5001  & 1.0002523 $\pm$ 0.0000048  & -77.5  & 2.0 \tabularnewline
{59 }  & C/2006 E1  & 0  & 5001  & 5001  & 1.000324 $\pm$ 0.000027  & -43.3  & 2.2 \tabularnewline
{60 }  & C/2006 K1  & 0  & 5001  & 5001  & 1.0024820 $\pm$ 0.0000088  & -352.4  & 0.9 \tabularnewline
{63 }  & C/2007 JA$_{21}$  & 0  & 5001  & 5001  & 1.000091 $\pm$ 0.000025  & -9.1  & 2.0 \tabularnewline
\hline
\end{tabular}
\end{table*}

Observed perihelion distance and ecliptic inclination distributions
of all investigated comets one can find in the first rows of Tables~\ref{tab:sample_q_dist}--\ref{tab:sample_i_dist}.
There is a clear observational selection signature in the perihelion
distance distribution. The ecliptic inclination distribution presented
in Table~\ref{tab:sample_i_dist} is shown for two groups of comets:
with $q<4.5$\,au and $q\geq$4.5\,au. In the whole sample of the
large perihelion Oort spike comets discovered since 1970 there are
more comets moving in prograde orbits (56.2 per cent) than moving
in retrograde orbits (43.8 per cent), however, for the sub-sample
of comets with 3.0\,au$\leq q<$4.5\,au we observe the same number
of prograde and retrograde orbit comets. It means that this disproportion
comes entirely from a subset of comets with q$\geq$4.5\,au, where
the ratio of comets in prograde orbits to comets in retrograde orbits
is about~1.6. (see Table~\ref{tab:sample_i_dist}).

We have also found an interesting feature in the
distributions presented in Table~\ref{tab:sample_i_dist}.  While the
number of prograde and retrograde orbits of comets with $3.0<q<4.5 $~au
is equal, the proportion of dynamically new to dynamically old ones
is reversed in these groups. In other words on prograde orbits there
are about twice as many dynamically new comets than on retrograde
orbits (9 : 4) and the proportion is opposite for the dynamically old 
comets (5:10).

A nice example of comet \emph{dynamically new} for sure, C/1999~J2,
is presented in Fig.~\ref{fig:gal_1999j2}. The previous perihelion
passage of nominal orbit of this comet happened some 9\,million years
ago at the heliocentric distance of 200\,au (80 per cent of its VCs
have a previous perihelion distance in the interval between 160 and
$\sim$250\,au, see Table~\ref{tab:past_motion_new}).

The spatial distribution of aphelion directions of all Oort spike
comets with $q>3$\,au (discovered before 2009) is presented in Fig.~\ref{fig:aitoff_galactic}.
Circles mark dynamically new comets, squares dynamically old ones
and triangles show seven comets with the uncertain dynamical status,
according to the definitions adopted in this section. In addition
to these we included here 11 large perihelion distance comets (marked
with crosses) discovered before 1970 and omitted in the present investigation.
Such a spatial distribution is often used when searching for a signature
of some specific perturbers, see \citet{matese_whitmire:2011} and
\citet{fernandez:2011} for recent discussions of this subject. When
looking for effects of a massive perturber moving on a distant heliocentric
orbit one should expect a concentration of the aphelia directions
along some great circle in the sphere. It seems to be difficult to
interpret the distribution presented in Fig.~\ref{fig:aitoff_galactic}
in such a way.

\subsection{How are they different from each other?}

\label{sub:Two_populations}

\begin{figure}
\begin{centering}
\includegraphics[width=8.6cm]{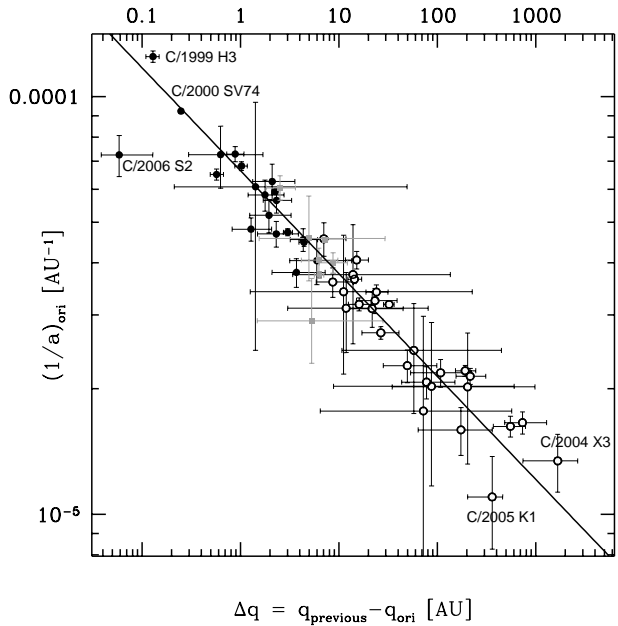} 
\par\end{centering}

\caption{$\Delta q$ vs $1/a_{{\rm ori}}$ in logarithmic scales, derived
for the observed sample of large perihelion Oort spike comets. The
open dots show the dynamically new comets, the filled dots -- the
dynamically old comets and grey squares - the seven comets with uncertain
dynamical status, i.e. comets not dynamically new for the limit of
15\,au for previous perihelion distance but being dynamically new
for the limit of 10\,au. The error bars for perihelion distances
are described by deciles within 10--90 per cent, the vertical error
bars are described by 1-sigma error for normal distribution of $1/a_{{\rm ori}}$.
Straight line represents the best fit to all presented points, except
the Comet C/2006~S2. }

\centering{}\label{fig:qpre_vs_a} 
\end{figure}

Figure~\ref{fig:distribution_a_all} shows the distribution of original
and future $1/a$ as well as the distribution of planetary perturbations
acting on comets during their passage through the inner solar system
($\Delta(1/a)=1/a_{{\rm fut}}-1/a_{{\rm ori}}$).
The black and white parts of histograms represent the dynamically
new and dynamically old comets, and grey -- comets with uncertain
dynamical status in the sense described above (Section \ref{sub:How-we-distinguish}).
Firstly, we focus on the total distributions of cometary energies
(left-hand side panels). It is clear that all three distributions
visible in Fig.~\ref{fig:distribution_a_all}\,(a) show some deviations
from the Gaussian model. However, the Gaussian fitting to the sample
of 62 comets (C/1978~G2 and C/1999~H3 were excluded) give $<1/a_{{\rm ori}}>=(37.8\pm17.7)\times10^{-6}$\,au$^{-1}$
with negative kurtosis%
\footnote{We use standard definition of kurtosis: $K=\frac{\mu_{4}}{\sigma^{4}}-3\,$,
where $\mu_{4}$ is the fourth central moment, $\sigma$ is the standard
deviation.%
} equal to -0.55. Generally $1/a_{{\rm ori}}$-distribution has
a wider and lower peak around the mean value than normally distributed
variable. Outside the horizontal scales of the middle and lowest panels
are two comets that have suffered large planetary perturbations during
their passage through the inner solar system: C/1980~E1~Bowell ($\Delta(1/a)=-16064\times10^{-6}$\,au$^{-1}$
mainly due to Jupiter encounter within 0.228\,au on December 1980)
and C/2002~A3~LINEAR ($\Delta(1/a)=+6153\times10^{-6}$\,au$^{-1}$
mainly due to Jupiter encounter within 0.502\,au on January 2003).
The planetary perturbations acting on the sample of these large perihelion
comets show clear asymmetry relative to zero (see lower panel) with
the negative median value of $\Delta(1/a)=-51.8\times10^{-6}$\,au$^{-1}$
whereas the distribution of $1/a_{{\rm fut}}$ are more symmetric
relative to zero (the middle panel) with median value of $-6\times10^{-6}$\,au$^{-1}$.
Statistical analysis shows that the $\Delta(1/a)$-distribution
is closest to the Gaussian distribution. We estimated the value of
mean planetary perturbations (represented by the standard deviation
of the $\Delta(1/a)$-distribution) equal to $285\times10^{-6}$\,au$^{-1}$
by fitting to the Gaussian distribution, however $\Delta(1/a)$-distribution
has the non-zero kurtosis (0.201) and is asymmetric with the longer
right tail (skewness equal to 0.237). The obtained mean planetary
perturbation in comet energy is significantly smaller than predicted
by the numerical simulations \citep{fernandez:1981,duncan-q-t:1987}.
This is probably the result of the non-uniform inclination distribution
in the observed sample of large perihelion comets (see Table \ref{tab:sample_i_dist}),
in contrast to the quoted simulations. With relatively large number
of high inclination and retrograde orbits the mean energy change is
expected to be smaller.

Future $1/a$-distribution seems to have the second small maximum
in the interval of $500\times10^{-6}$\,au$^{-1}\,<1/a_{{\rm fut}}<600\times10^{-6}$\,au$^{-1}$
(middle panel of Fig.~\ref{fig:distribution_a_all}). This maximum
consists of six comets (C/1974~F1, C/1974~V1, C/1992~J1, C/1993~K1,
C/1999~U1 and C/2000~CT$_{54}$) where three of them have $q_{{\rm osc}}<3.5$\,au.

Separate $1/a$-distributions of comets with moderately large $q$
(3.0\,$\leq q_{{\rm osc}}<4.5$\,au) and very large $q$ ($q_{{\rm
osc}}>4.5$\,au) are shown in panels (b) and (c) of  
Figure~\ref{fig:distribution_a_all} where the contributions of all
three dynamical groups of comets to original and future $1/{\rm
a}$-distributions as well as to $\Delta(1/a)$-distributions are
presented. Both distributions of original $1/a$ are noticeably
different. $1/a_{{\rm ori}}$-distribution of comets with very
large $q$ is more compact with a clear maximum around $1/a_{{\rm
ori}}\sim40-50\times10^{-6}$\,au$^{-1}$ whereas the distribution of
comets with moderately large $q$ is more flattened and is located
somewhere between $30-40\times10^{-6}$\,au$^{-1}$. An additional
difference is reflected in the absence of very large perihelion
comets with $1/a_{{\rm ori}}>70\times10^{-6}$\,au$^{-1}$ when we
have four such comets with moderate $q$. It may be noted that future
$1/a_{\rm ori}$-distribution and $\Delta(1/a_{\rm
ori})$-distribution for moderate $q$ comets are much more flattened
while for comets with very large $q$ we observe a clear maxima more
or less around zero. Thus, we observe the excess of comets with very
large $q$ experiencing small
($\mid\Delta(1/a_{{\rm ori}})\mid<10^{-4}$au$^{-1}$) planetary
perturbations.

Figure~\ref{fig:qpre_vs_a} shows the relation between the changes
in perihelion distance during the last orbital revolution and original
semimajor axes. Filled dots represent dynamically old comets, open
dots -- dynamically new comets and grey squares -- seven comets with
uncertain dynamical status. Six dynamically new comets with negative
$\Delta q=q_{{\rm prev}}-q_{{\rm ori}}$ and four dynamically new
comets with escaping or almost escaping swarm are not included in
this figure (see also Section~\ref{sub:Extremely-large-past-semimajor-axis})
except comet C/2005~K1 which is shown as lowest point in the figure.
Point standing off to the left of the fit line represents comet C/2006~S2.
It is worth noting that the comet C/1978~G2 with a formally negative
$1/a_{{\rm ori}}$ but large uncertainty of $1/a_{{\rm ori}}$-value
is consistent with the presented fit (the logarithmic scale in the
figure makes it impossible to show that). Straight line represents
the best fit to 53 points where the relation $\Delta q\sim\left(1/a_{{\rm ori}}\right)^{-4.06\pm0.16}$
was derived. The obtained exponent is significantly smaller than that
presented in \citet{yabushita:1989} but remarkably closer to the
expectations of the first order Galactic disk tide theory, \citep{byl:1986}.
It is still a little bit greater than theoretical but this might be
the result of including the Galactic centre tide in our model.

\begin{figure*}
\begin{centering}
\includegraphics[width=5.73cm]{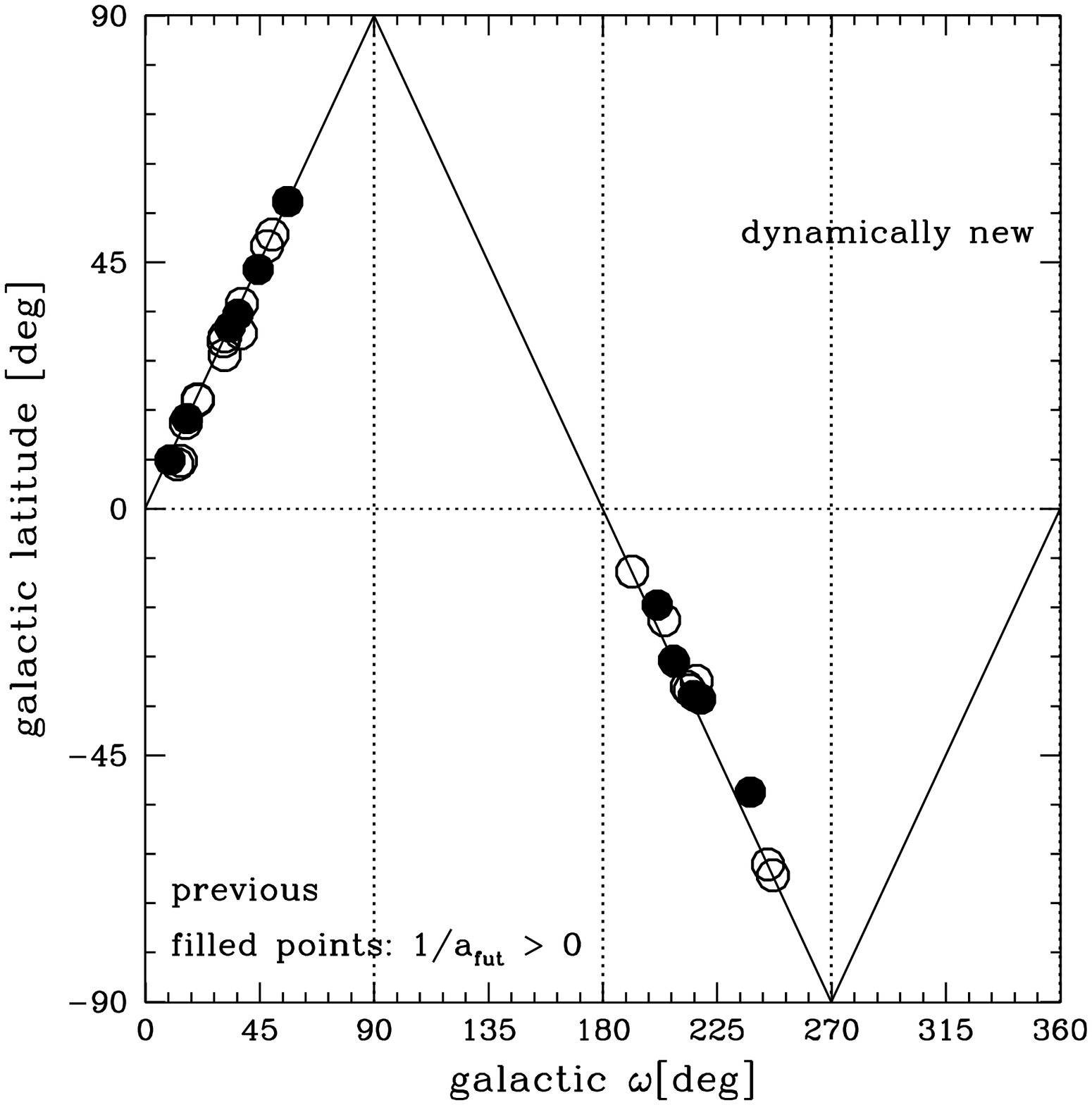} \includegraphics[width=5.73cm]{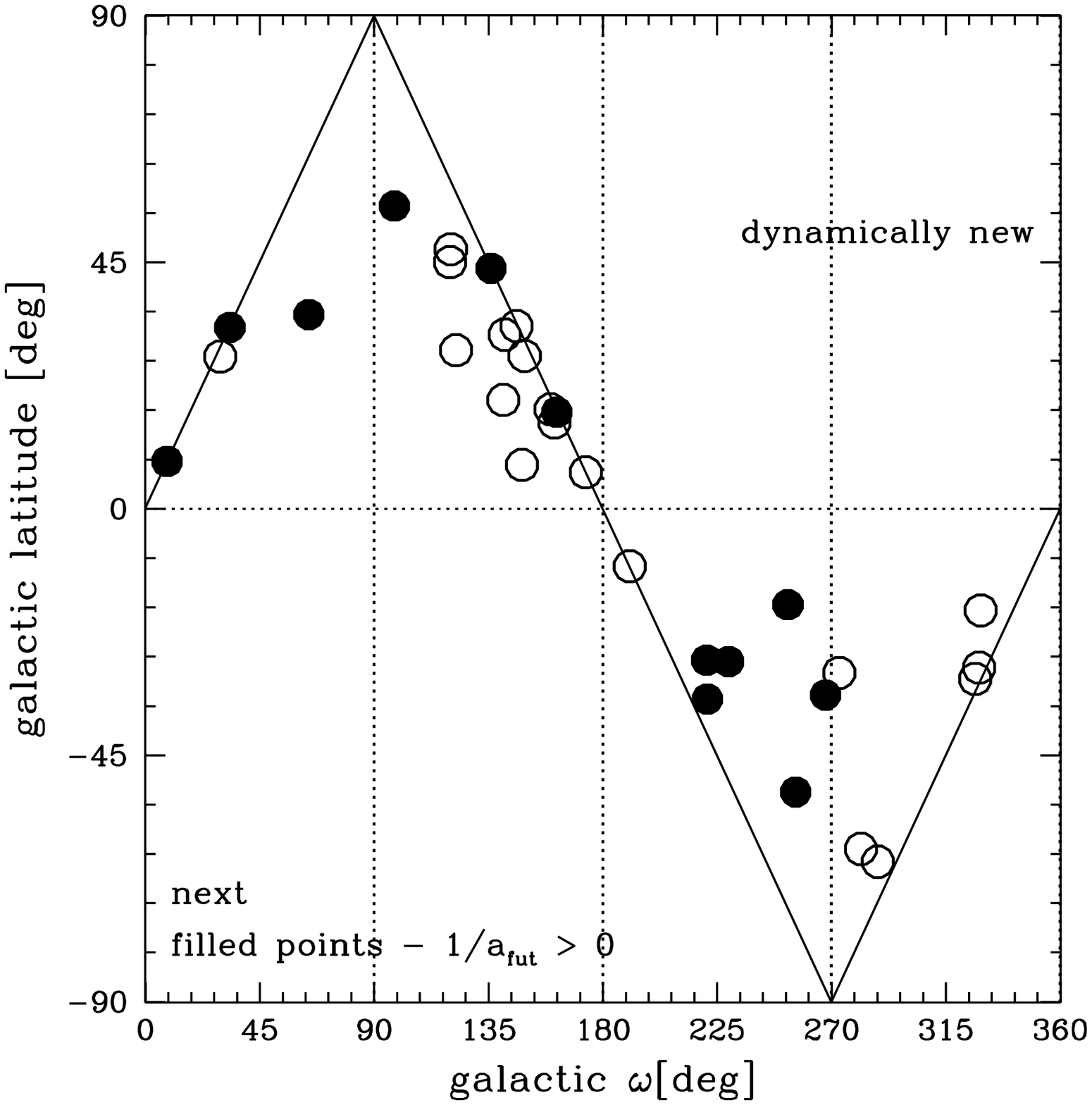}
\includegraphics[width=5.73cm]{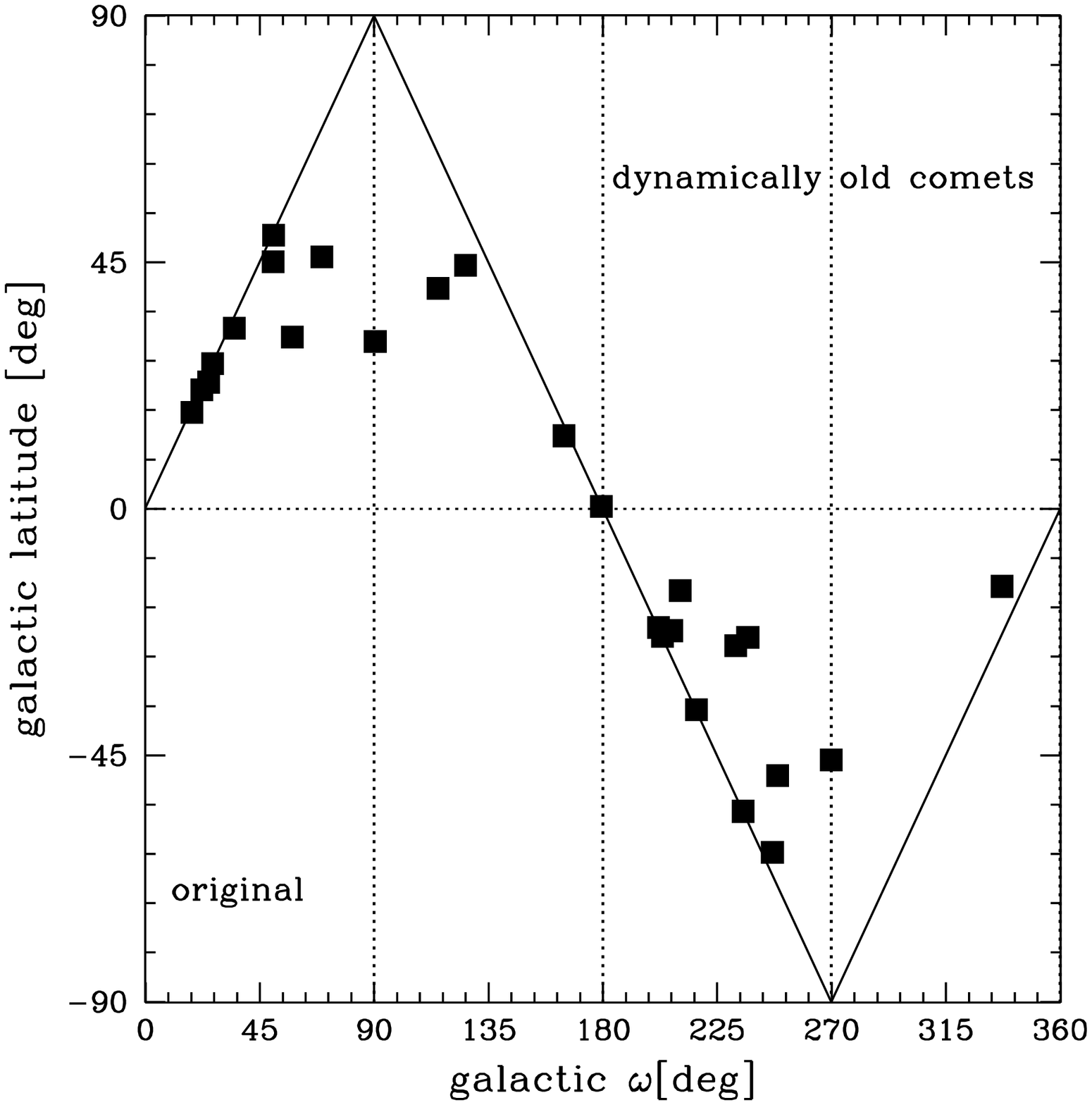} 

\par\end{centering}

\begin{centering}
(a) \hspace{6.0cm} (b) \hspace{6.0cm} (c) 
\par\end{centering}

\caption{Galactic $\omega$ vs galactic latitude for previous perihelion passage
or previous escaping on 120\,000\,au from the Sun (left-side and
right-side pictures for dynamically new and dynamically old comets,
respectively) or for the next perihelion passage or next escaping
on 120\,000\,au from the Sun (middle picture for dynamically new
comets). Right picture: almost all comets have galactic $\omega$
inside the first and third quarter; in the second and fourth quarter
are five of six comets with $q_{{\rm ori}}>q_{{\rm prev}}$: C/1972
L1, C/1976~D2, C/1980~E1, C/1997~J2, and C/1979~M3. }

\centering{}\label{fig:gal_pre} 
\end{figure*}

\subsection{Evolution of orbital elements in Galactic frame}

\label{sub:Evolution-of-orbital-elements-in-Galactic-frame}

It is a well known fact that under the separate Galactic disk tide
the secular evolution of cometary orbital elements is strictly periodic
and synchronous, i.e. the minimum of the perihelion distance coincides
with the minimum of the Galactic inclination, with the maximum of
the eccentricity and with Galactic argument of perihelion crossing
90 or 270 degrees. Including of the Galactic centre term introduces
only a small discrepancy from this regular patterns, but even during
one orbital period in can manifest itself in some specific cases.
This is true especially for small (Galactic) inclination orbits and
for large orbital periods.

\begin{figure*}
\begin{centering}
\includegraphics[width=8cm]{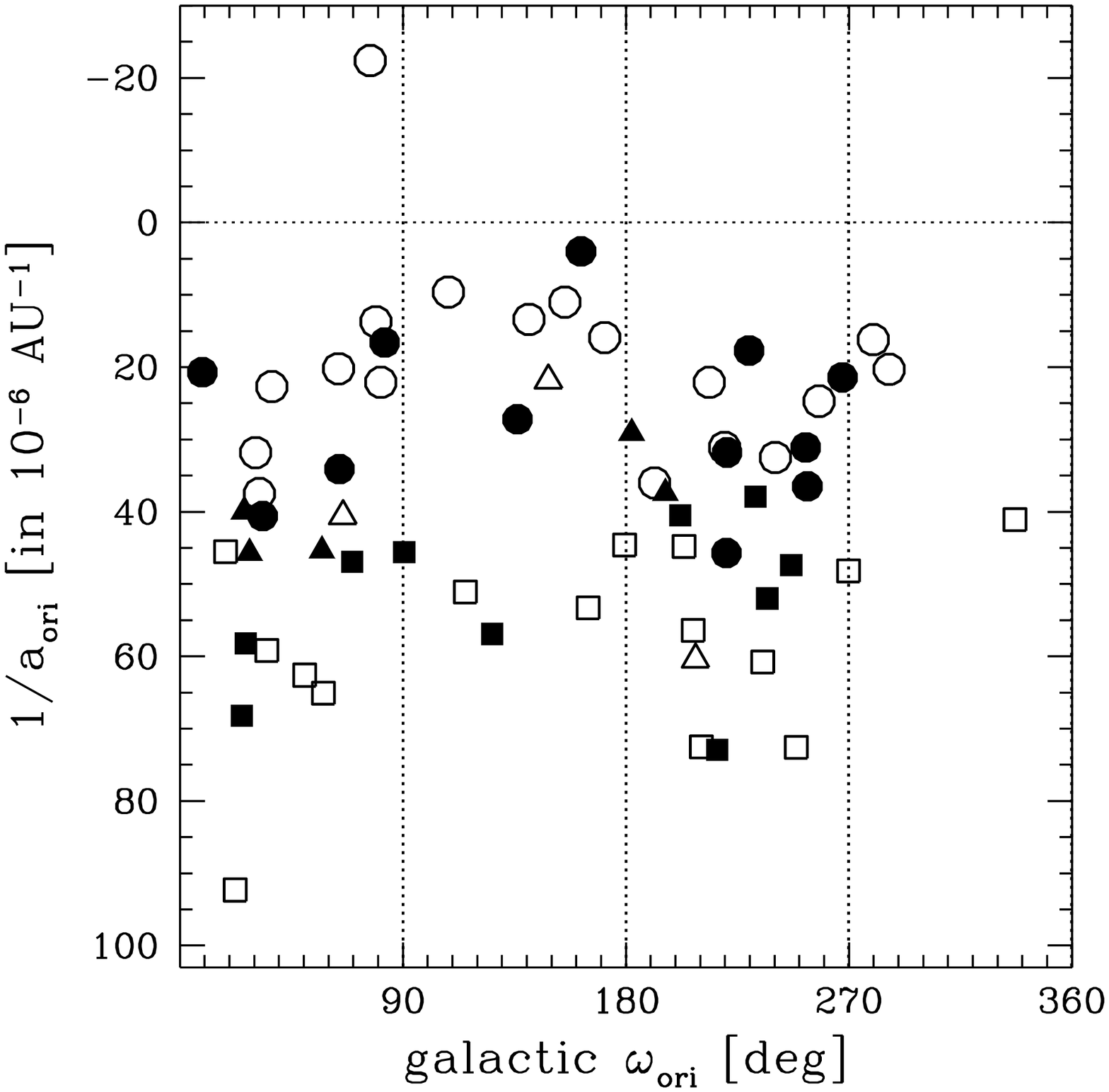} \hspace{1.0cm}\includegraphics[width=8cm]{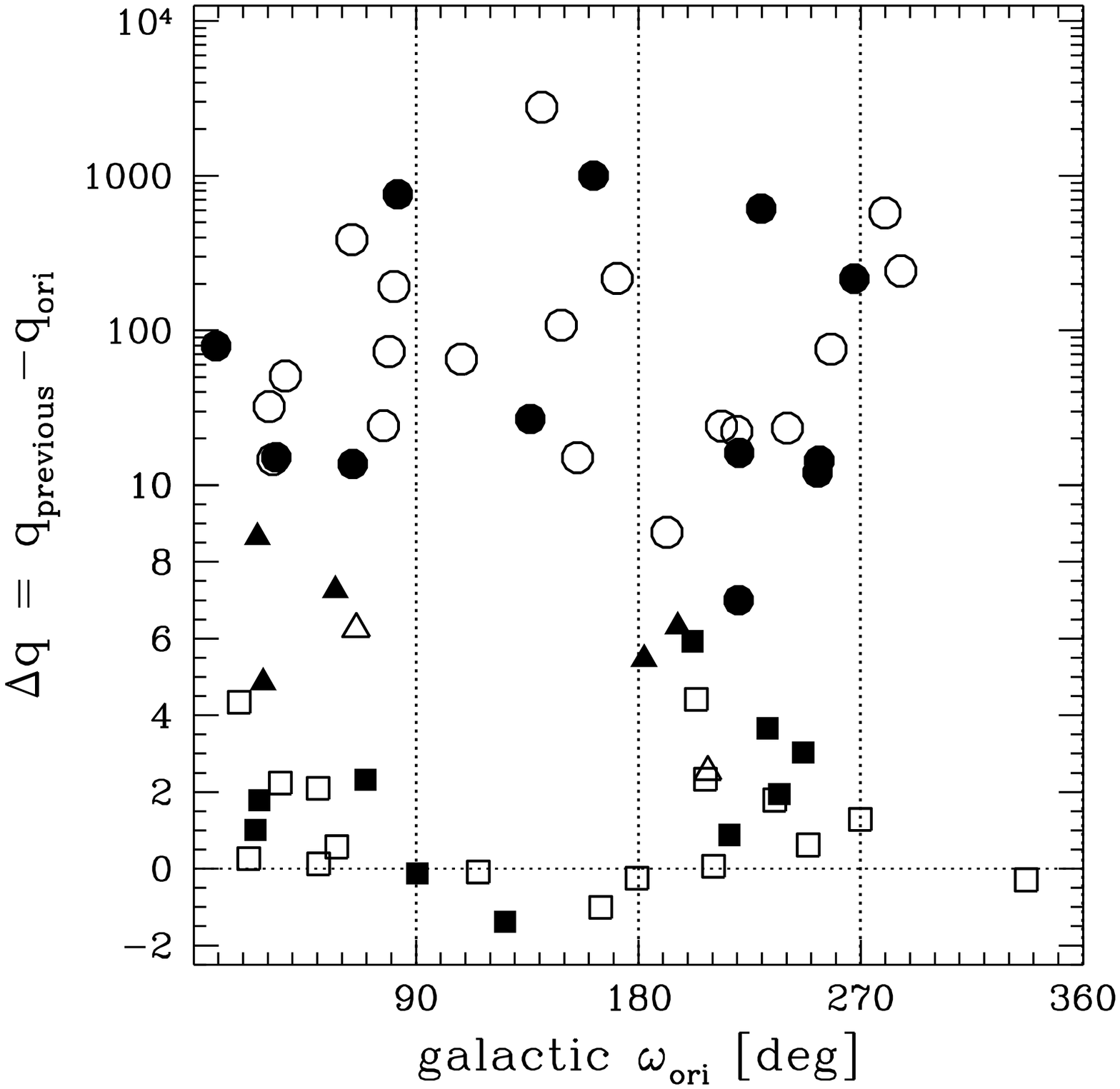} 
\par\end{centering}

\begin{centering}
(a) \hspace{8.0cm} (b) 
\par\end{centering}

\caption{Galactic argument of perihelion $\omega$ vs $1/a_{{\rm ori}}$ (left panel)
and vs $\Delta q=q_{{\rm prev}}-q_{{\rm ori}}$ (right panel) for all investigated
comets (note the change from the linear scale to a logarithmic one
in the middle of the vertical axis of the right panel). Circles and
squares represent dynamically new and dynamically old comets, respectively,
triangles -- comets with uncertain dynamical status. Filled and open
symbols mean returning or escaping in the future. Six comets with
the negative $\Delta q$, i.e. with $q_{{\rm ori}}>q_{{\rm prev}}$
(see also Table \ref{tab:q_increase}), can be located in the bottom
part of this figure. These are (from the left to right side): C/1976~U1,
C/1972~L1, C/1976~D2, C/1980~E1, C/1997~J2, and C/1979~M3.}

\centering{}\label{fig:gal_deltaq} 
\end{figure*}

Analysing the Galactic evolution of cometary perihelion distances
\citet{matese-lissauer:2004} introduced so called tidal characteristic
'$S$' which tell us whether the perihelion distance of a particular
comet decreases ($S=-1$) or increases ($S=+1$) during the observed
perihelion passage. In the dynamical model restricted to the Galactic
disk tide $S=-Sign(\sin(2\omega)$). Due to the relatively short time
intervals, limited to one orbital period we, despite of including
the Galactic centre term in our calculations, do not observe any departure
from this simple relation. It means, that for all comets investigated
here their perihelion distance decreased due to the combined Galactic
tides when Galactic argument of perihelion $\omega$ is in the first
or third quarter and increases otherwise. As discussed in \citet{matese-lissauer:2004}
and recently in \citet{matese_whitmire:2011} this is the reason that
dynamically new comets should have $S=-1$ (and $\omega$ is in the
first or third quarter) much more frequently. This, together with
the distribution of the Galactic latitude of perihelion direction
$b$ is illustrated in Fig.~\ref{fig:gal_pre}.

As in the previous plots, circles denotes here dynamically new comets
and squares dynamically old ones; in the left and middle panel filled
symbols mark comets returning in the future while open symbols mark
comets ejected in the future from the solar system as moving along
a hyperbolic orbits. Due to the relation $\sin b=\sin\omega\sin i$
comets may appear in this plots only between horizontal $b$-zero
axis and the solid, {}``triangle'' lines. All angles are measured
in the Galactic frame.

In the left panel of Fig.~\ref{fig:gal_pre} we present perihelion
latitude verse argument of perihelion for nominal orbits of dynamically
new comets at their previous perihelion passage. In accordance with
the patterns described above all circles are located along solid borders
of the allowed region, and only in the first and third quarter of
$\omega$ what corresponds to the perihelion distance decreasing phase
of its Galactic evolution ($S=-1$).

The same comets but at the next perihelion passage (i.e. two orbital
revolution later) are plotted in the middle panel of Fig.~\ref{fig:gal_pre}.
One can observe, that almost all comets moved significantly and all
four quarters are populated more or less uniformly. There is also
an intriguing asymmetry in the displacements of escaping and returning
comets. Please note, that the displacement of comets between the left
and the middle panel incorporates both Galactic orbital evolution
during two consecutive revolutions and planetary perturbations during
the observed perihelion passage.

The right panel of Fig.~\ref{fig:gal_pre} is to be compared with
the middle one. We present here angular elements of all (both returning
and escaping) dynamically old comets at their previous perihelion
passage. If our definition of being dynamically old or new is relevant,
new comets at their next perihelion ( only the filled points in the
middle panel ) should have angular elements distribution qualitatively
similar to that of the old comets at their original perihelion passage
(right panel). As one can easily observe we obtained a high level
of such a similarity.

\citet{matese-lissauer:2004} suggested that it should be a significant
correlation between the tidal characteristics $S$ ( i.e. the quarter
of $\omega$) and the reciprocal of the semimajor axis $1/a_{{\rm ori}}$.
As presented in Fig.~\ref{fig:gal_deltaq} much more clear correlation
can be observed between $\omega$ and the change in the perihelion
distance during the last orbital revolution: $\Delta q=q_{{\rm prev}}-q_{{\rm ori}}$.
It can be noticed that there are two distinct group of comets that
can be found in all four quarters of $\omega$: dynamically new comets
(that have largest $\Delta q$) and dynamically old comets with negative
$\Delta q$ (listed with details in Table~\ref{tab:q_increase}).
The rest of the dynamically old comets as well as all dynamically
uncertain ones can be found only in the first and the third quarter
of $\omega$.

\subsection{New interpretations}

\label{sub:New-interpretations}

Recently an interesting new scenario of the LPCs dynamical evolution
was pointed out by \citet{kaib-quinn:2009}. They showed, that in
addition to preventing some LPCs to be observed, the Jupiter-Saturn
barrier can help some inner Oort cloud comets ($a<10\,000$\,au)
to reach the observability zone. During the numerical simulation of
the Oort cloud dynamical evolution they observed, that large percentage
of the inner cloud comets follows a common pattern: due to weak Galactic
perturbations their perihelia slowly drift towards the Sun and at
heliocentric distances of 15--20~au they received a series of energy
kicks from the giant planets increasing significantly their semimajor
axis, see Fig.~1 in the quoted paper. At the late stage of this process
a comet with $q\simeq12$\,au goes outside the Planetary System along
the orbit with $a\simeq28\,000$\,au and then it returns to the Sun
at any\emph{ }small perihelion distance due to strong Galactic perturbations
received during this last orbital revolution.

\begin{figure}
\includegraphics[height=1\columnwidth,angle=-90]{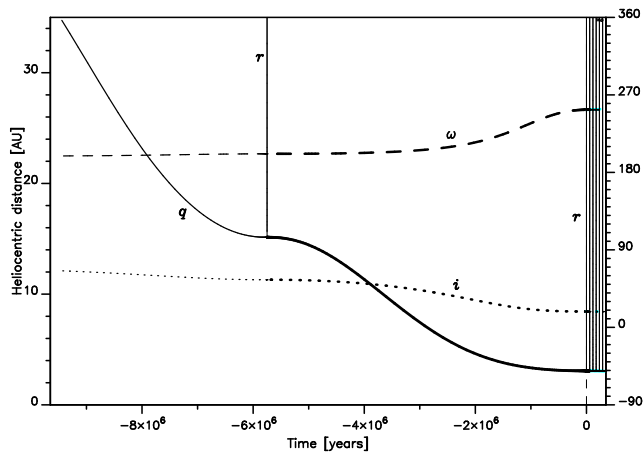}

\caption{\label{fig:gal_2001k3}Past and future orbital evolution of the nominal
VC for C/2001~K3. This comet came to the observability zone having
$a_{{\rm ori}}\sim32\,000$~au and might be considered as an example of
the probable output of the mechanism proposed recently by \citet{kaib-quinn:2009}.}

\end{figure}

They concluded that the majority of the observed LP~comets can be
produced by such a mechanism but except mentioning two objects in
strange orbits, 90377~Sedna \citep{brown_et_al:2004} and 2006~SQ$_{372}$
\citep{kaib:2009}, they do not provide any real cometary examples.

In our study we obtained several results which can fit their prediction
perfectly. By quick examination of Table~\ref{tab:past_motion_new}
one can find that C/1978~A1, C/1997~BA$_{6}$, C/2001~K3, C/2003~S3,
C/2004~T3, C/2007~Y1 are good candidates, as well as all 7 comets
from Table~\ref{tab:past_motion_uncertain}. Previous perihelia of
all these comets (as well as many VCs of other comets) lie in the
vicinity of the Jupiter-Saturn barrier and it seems to be quite achievable
that some of these comets were produced by the mechanism proposed
by \citet{kaib-quinn:2009}.

In Fig.~\ref{fig:gal_2001k3} we present the past and future Galactic
evolution of the nominal VCs of C/2001~K3. Its previous perihelion
distance and original semimajor axis fits perfectly to the scenario
proposed by Kaib and Quinn \citeyearpar{kaib-quinn:2009}. Visiting
the planetary system 5.7 million years ago at the heliocentric distance
of 11\,au it probably received significant perturbations from Jupiter
and Saturn. Therefore the probability that it was just placed on the
large semimajor orbit being previously the inner cloud member seems
to be quite significant. In the future this comet was captured into
small semimajor axis orbit of $\sim$1\,500\,au, leaving permanently
the Oort spike and having the next perihelion distance of equal to
the observed one (3.06\,au).

\section{Discussion and conclusions}

\label{sec:Conclusions}

As a continuation of Paper I we attempted to characterize past and
future motion of the next sample of the Oort spike ($1/a_{{\rm ori}}$<
1$\times10^{-4}$\,au$^{-1}$) comets. In order to minimize possible
biases due to indeterminable NG effects we decided to study here only
comets having perihelion distance $q_{{\rm osc}}>3$~au and precisely
determined orbits. To this aim we omitted both 11 LP comets with $q_{{\rm osc}}>3$\,au
but discovered before 1970 and three comets still (at the moment of
this writing) potentially observable. Such a complete sample of 64~large
perihelion distance Oort spike comets allows us to obtain some statistical
characteristics of this sample as well as individual past and future
dynamics for all of them. In the process of osculating orbit determination
we succeed with NG effects detection for 15~comets. Having a homogeneously
obtained set of osculating orbits we followed numerically their motion
among planets back and forth, up to the heliocentric distance of 250\,au,
obtaining original and future orbits. Instead of integrating one orbit
per comet we replaced each body with a set of 5001~VCs and followed
their motion individually. This allowed us to obtain all parameters
of the original and future orbits together with their uncertainties.

\begin{center}

\begin{table}
\caption{\label{tab:stellar-gal}The comparison of the previous perihelion
distance (all values are expressed in au) of C/1984 W2, C/1993 K1,
C/1997 A1 and C/1997~J2 obtained in the present paper and in \citet{dyb-hab3:2006}.
Demonstrated is both the influence of stellar perturbations and the
Galactic centre tide term.}

\begin{tabular}{llllcrr}
\hline 
\multicolumn{1}{l}{Comet} & \multicolumn{4}{c}{t~~~h~~~i~~~s \hspace{0.5cm} p~~~a~~~p~~~e~~~r} & \multicolumn{2}{c}{Dybczy{\'{n}}ski}\tabularnewline
\multicolumn{1}{l}{} &  &  &  &  & \multicolumn{2}{c}{(2006)}\tabularnewline
Name  & \multicolumn{1}{c}{$q_{{\rm ori}}$} & \multicolumn{1}{c}{$q_{d}$} & \multicolumn{1}{c}{$q_{dc}$} & \multicolumn{1}{c}{$q_{{\rm prev}}$ (all VCs)} & \multicolumn{1}{c}{$q_{d}^{*}$} & \multicolumn{1}{c}{$q_{ds}^{*}$}\tabularnewline
\hline 
C/1984~W2$^{NG}$  & 4.00  & 295  & 245  & 12.8 - 91.3 - 606  & ~~~~~490~~~~  & 571~~~~ \tabularnewline
C/1993~K1  & 4.85  & ~~~7.04 & ~10.3  & 6.33 - 10.1 - 32.9  & 6.29  & ~~7.08 \tabularnewline
C/1997~A1  & 3.16  & ~~78.4 & 111  & 56.7 - 111 - 225  & 124~~~~  & 141~~~~ \tabularnewline
C/1997~J2$^{NG}$  & 3.05  & ~~~2.99 & ~~~2.80  & 2.80 $\pm$0.02  & 2.96  & ~~2.59 \tabularnewline
\hline
\end{tabular}
\end{table}

\par\end{center}

Then we analysed past and future motion of all VCs for one
orbital period, including both Galactic disk and Galactic centre tides
and omitting the perturbations from all known stars. The later was
justified by \citet{dyb-hab3:2006}, who analysed by means of
the exact numerical integration the influence of all known stars on
the population of LPCs. In Table 6 of the quoted paper he listed
22 long period comets for which stellar perturbation changed the previous
perihelion distance by more than 10 per cent. Only four of the large
perihelion distance LPCs studied in the present paper can be found
in that table. Thus in Table \ref{tab:stellar-gal} we can compare those
results with the current calculations. There are several important
differences between calculations of \citet{dyb-hab3:2006} and the
present paper. First, we included here the Galactic centre tidal term,
noting its importance in some cases (\citet{dyb-hab3:2006} used only
the disc tidal term). Second, we homogeneously determined all cometary
orbits directly from observations and additionally included NG effects
where possible. Third, we replaced each comet with the swarm of 5001~VCs,
all compatible with the observations. This allowed us to observe the
influence of the propagated observational uncertainties on the final
results. 

In Table \ref{tab:stellar-gal} we present the current and
previous results for four comets: C/1984~W2, C/1993~K1, C/1997~A1
and C/1997~J2, two of them have detectable NG effects (marked by
the NG superscript after the comet designation). In the second column
we included the perihelion distance value for the original orbits
obtained in this paper. Next two columns show the previous perihelion
distance for the nominal orbit of these comets obtained without ($q_{d}$)
and with ($q_{dc})$ the Galactic centre tidal term. The next column
of Table \ref{tab:stellar-gal} , $q_{{\rm prev}}$, presents the result
of the observational uncertainties propagated back to the previous
perihelion. Last two columns show the previous perihelion distances
obtained by \citet{dyb-hab3:2006}: $q_{d}^{*}$presents the value
obtained only using galactic disk tide while $q_{ds}^{*}$denotes
the value obtained when perturbations of 21 most important stellar
perturbers were included. One can easily note, that the VCs previous
perihelion dispersion (column 5) is of one order greater that the
differences between current and previous results, except in the case
of comet C/1997~J2, where its current and past orbit can be determined
with the great accuracy. It can be observed that changes in the dynamical
model of the Galactic tides result in comparable (typically greater)
changes in the previous perihelion distance than after the incorporating
of the stellar perturbations (despite the fact, that these comets
are the most sensitive for known stellar perturbations, so these differences
should be treated as the extreme disparities). 

It should be noted, that even after including the action of
all 21 strongest stellar perturbers (still far to week to manifest
themselves) our classification of these four comets do not change
in any way: C/1984~W2 and C/1997~A1 are evidently dynamically new
comets, C/1997~J2 is evidently dynamically old one and C/1993~K1
remains the uncertain case. As it concerns the completeness of the
stellar perturbers search the reader is kindly directed to \citet{dyb-hab3:2006},
where possible sources of this incompleteness were discussed. We
only summarise here these arguments stating, that the omission of
any important (i.e. massive and/or slow and/or travelling very close
to the Sun) star seems rather improbable but not impossible. If such
a star will be discovered its dynamical influence on the long term
dynamical evolution of all LPCs should be carefully studied. This
would also may change some particular results and conclusions presented
here if it happened.

Additionally it seems worth to mention that the comet C/1997~J2 
belongs to an interesting group of six comets listed in Table \ref{tab:q_increase}, 
for which we obtained smaller previous perihelion distance than the 
observed one. As it is clear from Table \ref{tab:stellar-gal} none of the 
compared dynamical models do not change this classification also. 

The swarms of VC stopped at 250\,au from the Sun were next followed
numerically for one orbital revolution to the past and future and
as a result we obtained orbital characteristics of these objects at
the previous and next perihelion passages with respect to the observed
one. Both Galactic disk and Galactic centre tides were taken into
account for all comets. Basing on the previous perihelion
distance we divided a sample of 64 large perihelion Oort spike comets
into three groups: 26 dynamically old comets, 31 dynamically new comets
and 7 comets with the uncertain dynamical age. When analysing orbits
at the next perihelion we found that only 31 comets will remain Solar
System members in the future. The rest of our sample, 33 comets, will
leave our planetary system along hyperbolic orbits due to the planetary
perturbations.

Detailed results and plots for all individual comets studied in this
paper as well as summarizing catalogues of orbits will successively
appear at this project web page, \citet{dyb-krol:2011}.

As a result of studying all individual dynamical evolutions as well
as observing several interesting statistical characteristics of this
sample of comets we can draw following main conclusions: 
\begin{enumerate}
\item Observation selection and weighting are crucial for the precise orbit
determination. 
\item In contrary to popular opinion it is possible to determine orbits
with NG parameters for some 25 per cent of large perihelion distance
Oort spike comets. 
\item Incorporating these NG~orbits makes the overall energy distribution
of our sample significantly different, modifying the shape of the
Oort spike and position of the $1/a_{{\rm ori}}$-peak. 
\item Replacing each individual comet with a swarm of virtual comets {}``equally
well'' representing the observations is a powerful method for analysing
all uncertainties in the past and future motion of LP comets. This
allowed us for example to take the energy uncertainties into account
when preparing the energy distribution. 
\item In the absence of the recognized recent stellar perturbations it is
quite possible to calculate previous perihelion distances of the LP
comets, taking into account full Galactic potential. 
\item We have obtained clear correlation between the calculated change in
the perihelion distance and the reciprocal of original semimajor axis.
In contrast to some previous estimations we obtained the exponent
quite similar to the theoretically predicted. 
\item On the basis of the obtained previous perihelion distance distributions
we are able to distinguish comets that were observed for the first
time, i.e. dynamically new, from the rest of comets visiting the observational
zone at least twice -- these are called dynamically old. 
\item By defining three different threshold values for strong planetary
perturbations we can easily divide all 64 comets into 26 dynamically
new comets, 31 dynamically old comets and remaining 7 comets for which
their dynamical status seems to be uncertain. 
\item With the analysis of the Galactic angular orbital elements evolution
we found several significant fingerprints of the Galactic tides as
a dominating agent delivering observable Oort spike comets at present. 
\item In contrast to the overall picture of the so called Jupiter-Saturn
barrier we found that almost 50 per cent of our sample have the previous
perihelion distance below 15\,au. Moreover, we found several examples
of comets which have moved through the Jupiter-Saturn barrier almost
unperturbed. For six comets we found, that the observed perihelion
distance was even larger than the previous one. 
\item Among future orbits we found 27 comets that will be observable during
their next perihelion passage. In contrast, 33 comets will be lost
due to the hyperbolic ejection.
\item We have also found, that among 64~large perihelion distance Oort
spike comets almost 25 per cent can be treated as possible results
of the recently proposed by \citet{kaib-quinn:2009} new source pathway
from the inner Oort cloud to the observable zone. 
\end{enumerate}
\textbf{\emph{Acknowledgments.}} The research described here was partially
supported by Polish Ministry of Science and Higher Education funds,
(years 2008-2011, grant no. N N203 392734). Part of the calculation
was performed using the numerical orbital package developed by Professor
Grzegorz Sitarski and the Solar System Dynamics and Planetology group
at SRC. The authors wish also to thank Professor S\l{}awomir Breiter
for valuable discussions on some particular dynamical aspects of this
research. We also thank Professor Giovanni Valsecchi who,
as a referee, provided valuable comments and suggestions. This manuscript
was partially prepared with \LyX{}, the open source front-end to the
\TeX{} system.

\bibliographystyle{mn2e}
\bibliography{moja18}

\label{lastpage}
\end{document}